\documentclass[onecolumn,authoryear]{els-mrw} 

\usepackage{amsmath,amssymb,amsfonts,amsthm,makeidx,graphicx}
\usepackage{txfonts}
\usepackage{helvet}

\begin{document}

\chapter{Observations of the First Galaxies in the Era of JWST}\label{chap1}

\author[1]{Daniel P. Stark}%
\author[2]{Michael W. Topping}%

\author[3]{Ryan Endsley}%
\author[2]{Mengtao Tang}%

\address[1]{\orgname{University of California, Berkeley}, \orgdiv{Department of Astronomy}, \orgaddress{501 Campbell Hall, Berkeley, CA, 94720 USA}}
\address[2]{\orgname{Steward Observatory, University of Arizona}, \orgdiv{Department of Astronomy}, \orgaddress{933 N. Cherry Ave, Tucson, AZ 85721, USA}}
\address[2]{\orgname{University of Texas at Austin}, \orgdiv{Department of Astronomy}, \orgaddress{2515 Speedway, Austin, Texas 78712, USA}}

\articletag{Chapter Article tagline: update of previous edition, reprint}

\maketitle

\begin{abstract}[Abstract]
We provide a review of our current knowledge of galaxies throughout the first billion years of cosmic history. This field has undergone a transformation in the last two years following the launch of {\it JWST}, and we aim to deliver an observational overview of what we have learned about $z\gtrsim 5$ galaxies. We introduce the latest selection methods of high redshift galaxies and describe new measurements of the census of continuum-selected and dusty star forming galaxies at $z\gtrsim 5$. We discuss new measurements of the UV luminosity function at $z\gtrsim 10$ and associated implications for early star formation. We then summarize what  is being learned about the physical properties of early galaxies, with up-to-date discussions of the sizes, masses, ages, metallicities, abundance patterns,  UV colors, dust properties, and ionizing sources in $z\gtrsim 5$ galaxies.  We review  observational evidence for bursty star formation histories and describe prospects for characterizing the duty cycle with future observations. We provide a brief overview of the insight being gained through new detections of AGN in early galaxies. Finally we introduce the latest  constraints on the contribution of galaxies to reionziation and discuss how {\it JWST} measurements of Ly$\alpha$ emission offer the potential to probe the earliest stages of the process. This review is meant to provide a broad introduction to those new to the observational study of very high redshift galaxies. 
\end{abstract}

\section{Introduction}\label{sec:intro}

One of the major goals of observational astronomy is to stitch together a complete picture of cosmic history, from the Big Bang to the modern era. The cosmic microwave background (CMB) has long provided our first view, revealing the universe as it was just after recombination, roughly 400,000 years following the Big Bang. This epoch marks the beginning of the cosmic dark ages, when the universe was mostly permeated by neutral hydrogen, helium, and the other light elements formed after the Big Bang. Deep infrared images from large telescopes have long provided our next observational snapshot, revealing distinctly different 
conditions nearly one billion years later. Galaxies are common by this period, some already quite massive in stars ($\gtrsim 10^{10}$ M$_\odot$) and enriched in metals, others harboring supermassive black holes in their nuclei ($\simeq 10^{9}$ M$_\odot$), or tracing dense large-scale structures spanning over 1 physical Mpc. Deep quasar spectra reveal another major transformation: the neutral hydrogen (HI) that filled all of space after recombination becomes highly ionized throughout the intergalactic medium (IGM) by $z\simeq 5.3$, just over 1 billion years after the Big Bang.

Over the last few decades, theoretical work has constructed a framework that pieces together these radically different epochs. The dark ages are  terminated following the formation of the first stars from metal-free  gas clouds of primordial composition. These so-called Population III stars (Pop III) are expected to begin forming around $z\simeq 30$ in low mass minihalos that have collapsed and undergone H$_2$ cooling \citep[see][for a review]{Klessen2023}.  The initial mass function (IMF) of Pop III stars has long been a subject of debate. While our understanding is still quite limited, it is now thought that the Pop III mass spectrum spans a wide range, including both low  and high mass stars. The shape of the IMF is still  predicted to be  top-heavy, with relatively more massive stars per unit star formation than at the present day.  As the first supernovae explode, the  interstellar medium (ISM) becomes polluted with metals, eventually sparking a transition toward metal-enriched Population II (Pop II) stars. Once dark matter halos with virial temperatures above $\simeq 10^4$ K become common, efficient cooling via Ly$\alpha$ emission from atomic hydrogen can commence. It is these so-called atomic-cooling halos that host what are typically defined  as the first galaxies \citep{Bromm2011}. The massive stars that form in  the first stars and galaxies emit Lyman Continuum (LyC) radiation capable of ionizing hydrogen in the IGM.  As more galaxies emerge, intergalactic HII regions begin to emerge around galaxies, with the largest ``ionized bubbles'' expected to form in the most overdense regions of the universe. Eventually the bubbles overlap, and the process of hydrogen reionization is complete.

As  larger telescopes have emerged, the observational frontier has been steadily pushed back to earlier epochs, allowing this final missing chapter of history to be   filled in. Deep optical and infrared imaging with the {\it Hubble Space Telescope (HST)} drove progress for 25 years, providing the first census of star formation activity in the first billion years from statistical samples of color-selected galaxies at $4\lesssim z\lesssim 10$. Mid-infrared imaging from the {\it Spitzer Space Telescope} probed longer-wavelength light, constraining the build-up of stellar mass in early galaxies. The Atacama Large Millimeter Array (ALMA) has come to also play an important role, characterizing dust obscured star formation and ISM reservoirs in massive early galaxies. Ground-based  surveys targeting Ly$\alpha$ emission lines have begun to illuminate the timescale and morphology of reionization, complementing constraints from quasar spectra and the CMB. The most important recent developments have  come from the {\it James Webb Space Telescope (JWST)}. Its near-infrared camera has delivered unparalleled sensitivity at 1 to 5$\mu$m, extending the redshift frontier to $z\simeq 15$ and delivering new insight into the nature of the first generations of galaxies. The spectroscopic capabilities of {\it JWST} have proven revolutionary, opening up our first detailed window on the chemistry, stellar populations, and black holes of early galaxies, while also providing our first maps of large scale structure through the reionization era.

Given the recent  progress, it is a valuable time to update and review our current understanding of galaxies in the first billion years. While {\it JWST} will continue to deliver new results, much has been learned since the telescope launched and since the last observational reviews on the topic \citep{Dunlop2013,Stark2016, Robertson2022}. In this article, we aim to provide a timely reference for new researchers entering the field, highlighting the latest developments in our understanding of the first generations of galaxies. We will primarily focus on galaxies at $z\gtrsim 5$, but in places we will consider evolution in properties over a slightly larger redshift range. Our focus is primarily observational, and we direct readers to \citet{Dayal2018} for a more theoretically-based review.
The article is structured as follows. We describe galaxy selection methods (\S2) and provide an overview of the current census of high redshift galaxies in \S3. We then review what is now known about the star formation rates, stellar masses, and stellar population ages of the first generations of galaxies in \S\ref{sec:galaxyprops}, before discussing implications for star formation histories in \S5. The latest measurements of UV continuum slopes are presented in \S6, and sizes, morphology, and spatially-resolved properties are reviewed in \S7. Progress in our understanding of gas, dust, and metals is described in \S8. We discuss the nature of early ionizing sources in \S9 and the galaxy-halo connection and galaxy clustering in \S10. 
We provide a brief description of AGN in \S11, before discussing observational measurements of outflows in \S12. We close with a review of the contribution of galaxies to reionization and a discussion of  new measurements of Ly$\alpha$ emission at $z\gtrsim 6.5$  (\S12) before summarizing what has been learned in \S13.

\begin{figure}[t]
\centering
\includegraphics[width=.8\textwidth]{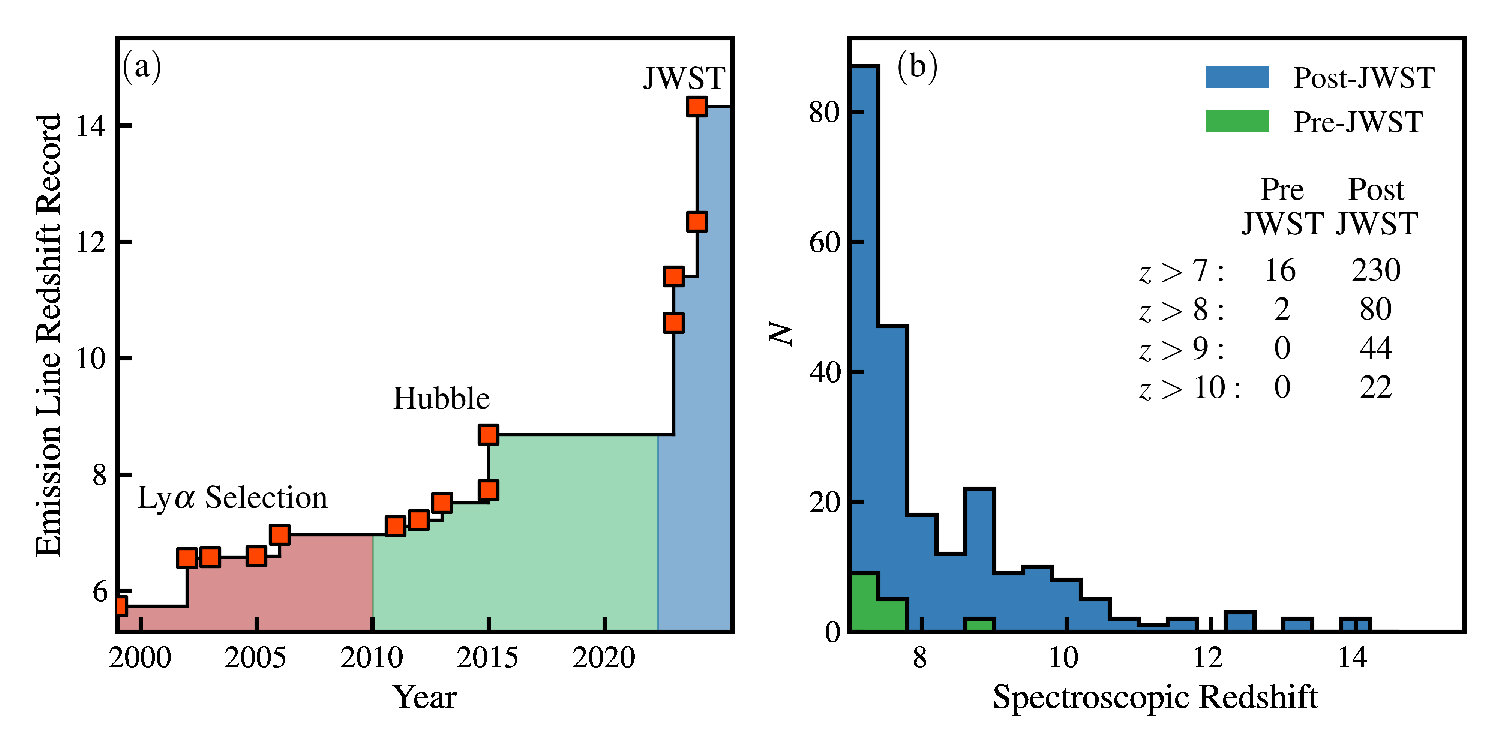}
\caption{{\it (Left):} Spectroscopic redshift record for galaxies since 2000. Narrowband Ly$\alpha$ selected galaxies were the most distant confirmed galaxies through 2011. Spectroscopic follow-up of continuum-selected  enabled a small number galaxies to be confirmed at $z=7-8.7$ between 2011 and 2022, with much progress coming from {\it WFC3}/IR imaging and new sensitive spectrographs. Since 2022, {\it JWST} has rapidly pushed back the redshift frontier, with the current record holder at $z=14.2$.
{\it (Right):} Histogram of the number of known $7<z<14$ spectroscopically-confirmed galaxies  before (green) and after (blue) the launch of {\it JWST.} Samples are from a recent compilation from M. Tang in deep fields with archival spectra. }
\label{fig:history}
\end{figure}

\section{Identifying Early Galaxies}\label{sec:earlygalaxies}
In this section, we review the basic methodology for selecting $z\gtrsim 5$ galaxies, focusing on continuum-selections (\S2.1), emission line selections (\S2.2), and identification of dust-obscured galaxies (\S2.3).
\subsection{Galaxy Continuum Selections}\label{sec:continuumselection}
Lyman Break Galaxy (LBG) selections have long provided the most efficient means of constructing large samples of high redshift galaxies. In absence of significant dust attenuation, the young stellar populations in  galaxies will power a strong rest-frame UV continuum spectrum.  At wavelengths below the Lyman limit ($\lambda_{\rm{rest}}$=912~\AA), the continuum will be heavily attenuated by neutral hydrogen in the galaxy and surrounding IGM. At $z\lesssim 4$, the emergent  spectrum of a galaxy will thus have a sharp spectral discontinuity (the so-called Lyman break) at the Lyman Limit, resulting in distinct  colors that allow for selection in multi-band imaging surveys.
The technique was first applied at $z\simeq 3$ over thirty years ago, using a combination of $U$, $G$, and $R$-band filters \citep{Steidel1992, Steidel1993, Steidel1995}.

Galaxies at higher redshifts ($z\gtrsim 4$) are easily identified by extensions of the LBG selection to redder observed-frame wavelengths. The denser IGM at $z\gtrsim 4$ provides stronger Lyman series opacity, causing the spectral discontinuity to shift from the Lyman limit to Ly$\alpha$ ($\lambda_{\rm{rest}}$=1216~\AA). 
A galaxy at $z\simeq 10$ ($\simeq 450$ million years after the Big Bang) will thus have a strong continuum break at $\lambda_{\rm{obs}}=1.3~\mu$m, between the $J$
and $H$-bands. Luminous galaxies at this redshift will appear bright and blue in $H$ and $K$-band filters and extremely faint in the J-band. Using simple color selections designed for specific redshift redshift ranges, it is possible build large samples of continuum-selected galaxies from  multiwavelength imaging. Photometric redshifts are also used to construct continuum-selected samples.  This approach has the benefit of using all available imaging data. In most cases at the redshifts considered in this review, the computed redshift will still depend most strongly on filters straddling the Ly$\alpha$ break, with  additional constraining power from filters sampling strong  emission lines in the rest-frame optical (i.e. H$\alpha$, [OIII]$\lambda\lambda$4959,5007).

The first statistical samples of continuum-selected galaxies at $z\gtrsim 6$ were assembled with imaging from the infrared channel on the Wide Field Camera 3 (WFC3/IR) of {\it HST}. WFC3/IR  provides photometric constraints 
over an 4.75 arcmin$^2$ field of view in four filters between the Y-band and H-band (F105W, F125W, F140W, F160W). When combined with deep optical imaging from the Advanced Camera for Surveys (ACS), the observed wavelength of the Ly$\alpha$ break can be constrained for galaxies spanning the redshift range $6\lesssim z \lesssim 10$. Following over 10 years of investment, WFC3/IR images reached depths of m$_{\rm{AB}}$=29--30, 27.5--27.8, and 26.6--26.8 in areas of roughly 4.7, 133, and 553 arcmin$^2$ \citep{Grogin2011, Koekemoer2011, Illingworth2013}, enabling identification of nearly 2000 galaxies with photometric redshifts at $z\simeq 6$, 7, and 8 \citep{McLure2013,Bouwens2015,Finkelstein2015}. 
At $z\gtrsim 9$, {\it HST} proved far less efficient owing in large part to the absence of imaging at wavelengths longer than the H-band. As the {\it JWST} era began in 2022, {\it HST} had only identified  roughly 40 photometric galaxies at $z\gtrsim 9$ \citep[e.g.,][]{Mcleod2016, Ishigaki2018, Morishita2018, Oesch2018, Bhatawdekar2019, Bouwens2021, Finkelstein2022a}.

The launch of {\it JWST} has quickly led to substantial improvements in galaxy continuum selections at the highest redshifts. NIRCam efficiently provides deep imaging in a suite of filters spanning 0.6--5$\mu$m, delivering the redder wavelength coverage required to more confidently identify the Ly$\alpha$ break at $z\gtrsim 9$. NIRCam provides simultaneous imaging over a 9.7 arcmin$^2$ field of view in its short wavelength (SW)  and long wavelength (LW) channel.  In its first two years of operations, NIRCam revisited many of the deep fields previously targeted by {\it HST} (GOODS-N, GOODS-S, UDS, COSMOS, EGS, Abell 2744), building up an area of 720 arcmin$^2$ with multi-band imaging reaching m$_{\rm{AB}}$=29.0--30.5 in the key 1--3 $\mu$m filters needed for identification of $z\gtrsim 9$ galaxies (F090W, F115W, F150W, F200W; \citealt{Treu2022, Finkelstein2023, Eisenstein2023,Bagley2024,Bezanson2024,Donnan2024}).  The longer wavelength  (3--5$\mu$m) filters used in these fields (F277W, F356W, F410M, F444W)
sample the spectrum at a variety of rest-frame wavelengths throughout the near-UV and optical, improving constraints on the  photometric redshifts. NIRCam imaging has also been conducted over a wide area (0.6 deg$^2$) in the COSMOS field with moderate depth (m$_{\rm{AB}}$=28) and four filters (F115W, F150W, F277W, F444W; \citealt{Casey2023}). Spectroscopic follow-up of NIRCam-identified galaxies quickly revolutionized the number of $z\gtrsim 7$ galaxies with confirmed redshifts (Fig 1b), extending the emission line redshift record to $z=14.2$ (Fig 1a). 
At slightly lower redshifts ($z\simeq 4-6$), galaxy selections still rely on {\it HST} imaging, but photometric redshifts are significantly improved thanks to NIRCam imaging of the rest-frame optical.

Over the last decade, several ground-based telescopes (e.g., Subaru, VISTA, UKIRT) have conducted very wide area ($>$1 deg$^2$) optical and infrared imaging surveys, complementing the smaller areas of the sky that {\it HST} and now {\it JWST} have surveyed. The ground-based campaigns reach much shallower depths ($J=25-26$) but are critical for identifying rare, luminous galaxies at the bright end of the galaxy population \citep[e.g.,][]{Stefanon2017, Ono2018, Bowler2020}. These efforts are particularly crucial for studies of reionization, given the wide areas ($>10$ arcmin$^2$) spanned by large bubbles in the latter stages of the process.
We are now entering a renaissance in wide-field imaging capabilities with the {\it Euclid Space Telescope} now delivering deep infrared imaging over extremely wide areas, and the {\it Roman Space Telescope} set to begin its mission in 2027. These surveys promise to make significant impact on our understanding of reionization and the most luminous early galaxies in the near future. 

\subsection{Emission Line Selections}\label{sec:emlineselection}
High redshift galaxies can also be selected based on the presence of  emission lines. With optical and infrared telescopes, emission line selections generally focus on either 
Ly$\alpha$ in the rest-frame UV or [OIII] and H$\alpha$ in the rest-optical. At $z\gtrsim 4$, the strong rest-optical lines are redshifted out of the K-band making them impossible to detect from the ground. 
Historically, emission line selections at these redshifts have instead focused on targeting the Ly$\alpha$ emission line, as it remains within the wavelength range probed by ground-based facilities out to very high redshifts ($z\gtrsim 10$). 
The standard method for identifying statistical samples of Ly$\alpha$ emitters utilizes  narrowband and broadband filters on imaging cameras. When strong emission lines are situated in narrowband filters, galaxies will appear extremely bright relative to adjacent filters which only sample continuum emission. Via these narrowband flux excesses, emission line galaxies can be efficiently identified. 
Traditionally Ly$\alpha$ searches have been conducted at specific redshift ranges ($z\simeq 5.7$, $z\simeq 6.6$, $z\simeq 7.1$) corresponding to observed wavelengths where Ly$\alpha$ sits in clean regions of the atmosphere \citep{Malhotra2004,Hu2010,Ouchi2010}. Simple selections can be applied to confirm narrowband excess sources as Ly$\alpha$ emitters. For the first decade of the 2000s, the narrowband Ly$\alpha$ selections provided our highest redshift confirmed galaxies (Fig. 1a). The narrowband Ly$\alpha$ emitters have been reviewed recently \citep[][]{Ouchi2020}, and we direct a reader there for a more detailed treatment. 

Recently attention has come to focus on rest-frame optical emission line selections at $z\gtrsim 6$, now uniquely possible given the sensitive 2--5$\mu$m capabilities of {\it JWST}. The method receiving the most attention thus far 
involves the NIRCam grism, which provides moderate resolution (R=1600) slitless spectroscopic capabilities in several filters at $2.4-5~\mu$m \citep{Rieke2023}. With its longest-wavelength filter (F444W), the NIRCam grism probes [OIII] emitters over $6.7\lesssim z \lesssim 8.9$. In the first cycle of {\it JWST} operations, grism observations were obtained in F444W over the two GOODS fields, building an area of over $\simeq 140$ arcmin$^2$ with typical line flux limits (5$\sigma$) of 2$\times$10$^{-18}$ erg cm$^{-2}$ s$^{-1}$ from 2 hour exposures \citep[e.g.,][]{Oesch2023}. 
Several other surveys have now obtained comparably deep grism exposures through broad and medium-bands at 3--5$\mu$m in a variety of extragalactic fields, including along $z\gtrsim6$ quasar sightlines and in lensing cluster fields \citep[e.g.,][]{Kashino2023,Li2023_magnif,Wang2023_ASPIRE,Naidu2024}.
By blindly providing spectroscopic redshifts for all line emitters in a given field, the grism provides the most efficient path to mapping large scale structures, identifying overdensities and voids in the galaxy distribution. This is important for identifying and characterizing regions that are likely to have carved out ionized bubbles, enabling reionization to be studied on local scales (\S13.2).

\subsection{Dust Continuum Selections}\label{sec:dustselection}

For more than a decade, we have known that some very high-redshift galaxies are heavily obscured by dust, causing the bulk of their far-UV and optical light to be reprocessed into far-infrared continuum emission \citep[e.g.,][]{Capak2011,Walter2012}.
These early dust-enshrouded galaxies can be efficiently selected using far-infrared (FIR) data that constrain the peak wavelength of the dust continuum SED, since galaxies at higher redshifts will have their peak displaced to longer wavelengths.
For a nominal dust blackbody temperature of $T\approx30-40$ K \citep[e.g.,][]{Schreiber2018}, the SED peak is located at $\sim$250$\mu$m at $z=2$ while it is displaced to $\sim$600$\mu$m at $z=6$.
By the early 2010s, \textit{Herschel} had delivered photometry at 250$\mu$m, 350$\mu$m, and 500$\mu$m across a large area (21 deg$^2$), enabling the selection of objects with SEDs that continued to rise to $\gtrsim$500$\mu$m \citep[e.g.,][]{Riechers2013,Dowell2014}.
This ultimately led to the discovery of the first dusty star forming galaxy (DSFG) at $z>6$, HFLS 3 at $z_\mathrm{spec}=6.3369$, with an enormous star formation rate of $\approx$3000 $M_\odot$/yr \citep{Riechers2013}.

Candidate high-redshift DSFGs can also be identified by searching for continuum sources in longer-wavelength data ($\approx$2--3mm), pairing flux constraints at slightly shorter wavelengths to rule out the most obvious low-redshift systems (see, e.g., \citealt{Strandet2016,Casey2018}).
Such 2--3mm surveys are proving to be relatively efficient at selecting $z>4$ DSFGs with obscured SFRs$\sim$500--1000 $M_\odot$/yr over a few hundred square arcminutes \citep[e.g.,][]{Magnelli2019,Casey2021,Long2024}, delivering a more complete census of the obscured star formation rate density within the first 1.5 billion years (see \S\ref{sec:dustSFRD}).

Prior to the launch of {\it JWST}, red-continuum sources were being identified from small-area ($\sim$10--100 arcmin$^2$) \textit{Spitzer}/IRAC and Atacama Large Millimeter Array (ALMA) maps that went undetected in deep {\it HST} imaging \citep[e.g.,][]{Huang2011,Franco2018,Williams2019,Wang2019,Sun2021}.
From their UV through far-infrared SEDs, many of these so-called `{\it HST}-dark' objects were thought to lie at $z\gtrsim4$ and have modestly large obscured star formation rates ($\sim$100 $M_\odot$/yr).
Indeed, a few systems were confirmed to lie at such redshifts from ALMA line detections \citep[e.g.,][]{Umehata2020,Fudamoto2021}.
The arrival of very deep near-infrared {\it JWST}/NIRCam imaging quickly delivered a much larger sample of very dusty high-redshift systems with SFRs clearly far below typical sub-mm selected DSFGs \citep[e.g.,][]{Barrufet2023,Endsley2023,PerezGonzalez2023,Rodighiero2023,Williams2024}.
At least a subset of these objects appear to harbor bright AGN (see \S\ref{sec:AGN}) and their contribution to the obscured star formation rate density remains under investigation. 

\section{Census of Galaxies in the First Billion Years}\label{sec:census}

In this section, we will describe efforts to measure the luminosity function of continuum-selected galaxies (\S\ref{sec:lf}) and discuss the number density of dusty star forming galaxies and their contribution to the star formation rate density of the universe (\S\ref{sec:dustSFRD}). Stellar mass functions are now starting to appear with NIRCam-based measurements. 
We will briefly introduce these in \S\ref{sec:dustSFRD} in the context of the contribution of dusty galaxies to the total mass budget at $z\gtrsim 5$. 

\begin{figure}[t]
\centering
\includegraphics[width=1\textwidth]{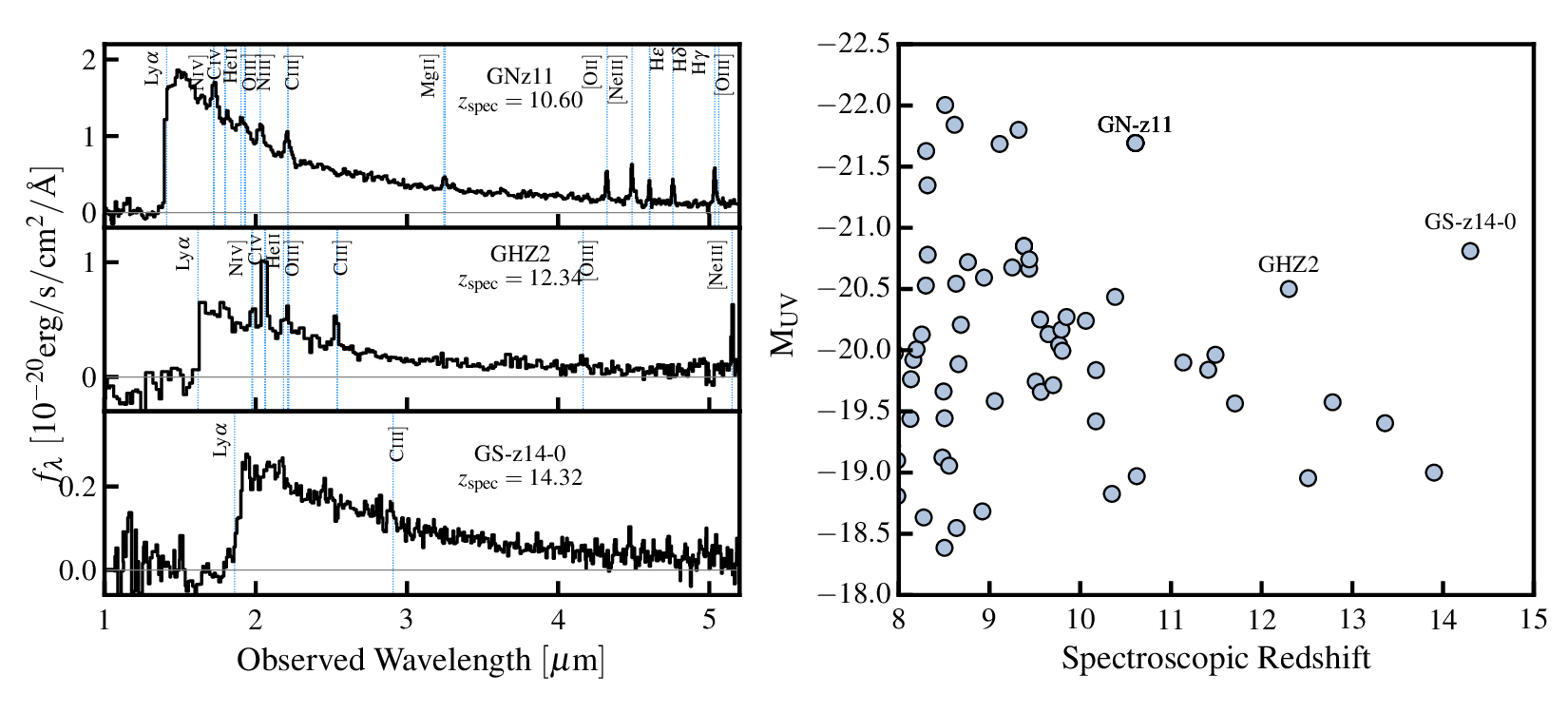}
\caption{(Left:) UV luminosities of spectroscopically-confirmed galaxies at $z>8$. Extremely luminous galaxies remain out to $z=14$.  (Right:) NIRSpec prism spectra of GNz11 \citep{Bunker2023}, GHZ2 \citep{Castellano2024}, GS-z14-0 \citep{Carniani2024}, the three brightest  galaxies known at $z>10.$ Two of the three (GHZ2 and GNz11) show intense emission line spectra that reveal hard ionizing sources. GS-z14-0 shows a much weaker emission line spectrum for unknown reasons.}
\label{fig:highzspectra}
\end{figure}

\subsection{Luminosity Function of Continuum-Selected Galaxies at $z>4$}\label{sec:lf}

The UV luminosity function (UVLF) of galaxies is often parameterized as a Schechter function \citep{Schechter1976} with the following form,
\begin{equation}
\phi(L) = \phi^\star \left( \frac{L}{L^\star} \right)^\alpha e^{-L/L^\star} 
\end{equation}
where $\phi^\star$ is the characteristic volume density, L$_{\rm{UV}}^\star$ is the characteristic UV luminosity (typically defined at 1500~\AA), and $\alpha$ is the faint end slope. This functional form is a good representation of galaxy populations with a power-law slope $\alpha$  extending to low luminosities and an exponential cutoff for luminosities greater than the characteristic luminosity L$^\star$.
For high redshift galaxies, the UV luminosity function is typically reported as a function of absolute magnitude (M$_{\rm{UV}}$), with the following form easily derived by converting L$_{\rm{UV}}$ to M$_{\rm{UV}}$.
\begin{equation}
\phi(\rm M_{UV}) = 0.4\ln(10)\phi^\star \left( 10^{-0.4(\alpha+1)(\rm M_{UV}-M_{UV}^{\star})} \right) exp\left[ -10^{0.4(\rm M_{UV}-M_{UV}^{\star})}\right] 
\end{equation}
where the UV absolute magnitude  is generally reported at a rest-frame wavelength of 1500~\AA. In most recent papers, M$_{\rm{UV}}$ is  derived by fitting stellar population models to the observed photometry and computing the absolute magnitude of the best-fit model at rest-frame wavelength of 1500~\AA. Such spectral energy distribution (SED) models will be described more in \S\ref{sec:galaxyprops}.

The {\it HST} imaging surveys described in \S\ref{sec:earlygalaxies} provided a census of UV continuum-selected galaxies at $z\simeq 4$ to $z\simeq 8$, constraining the volume density of UV-emitting galaxies with absolute magnitudes ranging from very faint (M$_{\rm{UV}}$=-17) to very bright (M$_{\rm{UV}}$=-21). 
The {\it HST}-based luminosity functions at $z\simeq 4-8$ have been presented by a variety of teams \citep[]{McLure2013,Finkelstein2015, Bowler2015, Ishigaki2018, Bouwens2021}, with a broad consensus indicating fairly smooth evolution at $z\gtrsim 4$.  Luminous galaxies (M$_{\rm{UV}}=-21$) are found to decrease in volume density by a factor of 19 in the 900 Myr between $z\simeq 4$ and $z\simeq 8$. The evolution is somewhat less rapid at the faint end of the luminosity function, with only a factor of 2.7 drop in the volume density of M$_{\rm{UV}}=-17$ galaxies between $z\simeq 4$ and $z\simeq 8$. When fit with a Schechter function, the redshift evolution of {\it HST}-selected samples can be explained by a steady decrease in the normalization parameter $\phi^\star$ coupled with a steepening of the faint end slope $\alpha$ toward higher redshifts, with little evolution in the characteristic luminosity, M$^\star_{\rm{UV}}$. The faint end slope is found to evolve from $\alpha=-1.7$ at $z\simeq 3$ to $\alpha = -2.0$ at $z\simeq 7$ \citep[e.g.,][]{Bouwens2015, Finkelstein2015, Bowler2020}, consistent with theoretical expectations \citep{Jaacks2012, Tacchella2013}.  We note that the luminosity function is expected to deviate from the faint end power law slope at some lower luminosity threshold. We will discuss current evidence for this low luminosity turnover at the end of this section.

By integrating the UV luminosity function, the UV luminosity density of the galaxy population ($\rho_{\rm{UV}}$) can be computed as a function of redshift.  Because of the steep faint end slope, the majority of $\rho_{\rm{UV}}$ will be emitted by faint galaxies.  When computing the evolution in $\rho_{\rm{UV}}$, it is common to adopt a lower bound on the UV luminosity that is  consistent with current observational limits (often M$_{\rm{UV}}$=-17), as this will minimize uncertainty in the location of the turnover in the luminosity function.   Simple conversions between SFR and L$_{\rm{UV}}$ allow $\rho_{\rm{UV}}$ to be converted to $\rho_{\rm{SFR}}$, the cosmic SFR density\footnote{The (dust-uncorrected) SFR is often derived from the emergent UV luminosity using the calibration adopted in \citet{Madau2014}, SFR$_{\rm{UV}}$ (M$_\odot$/yr) = 1.15$\times$10$^{-28}$ L$_{\rm{UV}}$ (erg s$^{-1}$ Hz$^{-1}$), assuming a Salpeter IMF. The precise calibration depends somewhat on stellar population ages and metallicity. The inferred SFR is then corrected for dust,  assuming a relationship between UV slope and M$_{\rm{UV}}$ (see \S\ref{sec:uvslopes}) and a relationship between  attenuation and UV slope (see \S\ref{sec:dustprops}, e.g., \citealt{Meurer1999}) }.  Prior to {\it JWST}, it was  shown that the integrated cosmic SFR density evolution at $5<z<9$  could be explained with  models in which the  star formation efficiency (defined here as the ratio of the SFR and halo mass growth rate) is constant with redshift \citep[e.g.,][]{Mason2015, Tacchella2018,  Bouwens2021, Harikane2022}, consistent with inferences from clustering measurements (see \S\ref{sec:clustering}, \citealt{Harikane2022}). The star formation efficiency (SFE) is found to depend on halo mass (M$_{\rm{halo}}$), with lower SFE generally expected in lower mass halos. In the `constant SFE' models used prior to {\it JWST}, the  derived relation between SFE and halo mass  was calibrated at $z\simeq 4-6$ using abundance matching and clustering, with very small (or no) scatter assumed in the SFE at fixed mass. 
Assuming a redshift-invariant SFE-M$_{\rm{halo}}$ relation, galaxy luminosity functions can  be predicted at yet higher redshifts ($z\gtrsim 10$), accounting for evolution in the underlying halo mass function and halo mass growth rates. The pre-{\it JWST} constant SFE models found excellent agreement with the luminosity function evolution at $z<10$, predicting that 
the number density of luminous galaxies steadily drops and faint end slope steepens toward higher redshifts, with the bulk of  star formation shifting to somewhat lower mass halos where the SFE is marginally lower.

Wide-area surveys conducted with ground-based telescopes have enabled the luminosity function measurements to be extended to brighter galaxies, most of which are too rare to be found in {\it HST}-based imaging surveys. These surveys have revealed a large population of extremely luminous  (M$_{\rm{UV}}=-23$) galaxies.  No strong evolution  
in the abundance of such luminous galaxies is found between $z\simeq 5$ and $z\simeq 9$ \citep{Bowler2020}, in contrast to the rapid drop in number density of galaxies with 
M$_{\rm{UV}}=-21$ over the same redshift interval. This 
is challenging to reproduce with a Schechter function, 
motivating use of a double power law (DPL) function for characterization of the early galaxy luminosity function,
\begin{equation}
\phi(M) = \frac{\phi^\star}{10^{0.4(\alpha+1)(M-M^\star)}+10^{0.4(\beta+1)(M-M^\star)}},
\end{equation}
where $\beta$ and $\alpha$ are the bright and faint-end slopes. The DPL function fits the luminosity function as well as the Schechter function at $z\simeq 5-6$ and is a much better fit  at $z\gtrsim 6$ \citep{Bowler2015, Bowler2020}. These results have suggested a departure from intermediate redshifts ($2\lesssim z\lesssim 5$) where a Schechter function is a better description of the luminosity function. It has been suggested that this may reflect the growth of dust and  onset of mass quenching in the galaxies occupying the extremely bright end of the luminosity function, both of which would act to reduce the number density 
of the brightest sources. In the context of the DPL formalism, the evolving luminosity function is driven primarily by evolution in M$^\star$. We provide best-fit parameters of DPL function fits to the luminosity function in Table~\ref{tab:uvlf}. 

{\it JWST} has made great strides in characterizing the luminosity function at $z\gtrsim 9$.  {\it HST}-based samples were too small at these high redshifts (\S\ref{sec:continuumselection}) for a robust picture. According to the constant SFE  models that reproduced the $4<z<8$ evolution, galaxies with M$_{\rm{UV}}<-17$ should be very rare at $z\gtrsim 10$.  The global SFR density (M$_{\rm{UV}}<-17$) was predicted to fall off very rapidly toward higher redshifts, as 
$\rho_{\rm{SFR}}\propto10^{-0.5(1+z)}$ (\citealt{Harikane2022}, see also \citealt{Mason2015, Mashian2016, Sun2016, Tacchella2018, Oesch2018, Behroozi2020}). This suggests that we should have seen a $\sim$1000$\times$ drop in the SFR density between $z\simeq 7$ and $z\simeq 14$, with UV bright galaxies (M$_{\rm{UV}}<-20$) becoming exceedingly rare at $z\gtrsim 10$.
 This is not what has been seen with {\it JWST}.  Deep NIRCam imaging revealed several bright $z\gtrsim 10$ galaxies in the first months of science operations \citep[e.g.,][]{Castellano2022, Donnan2023, Finkelstein2022, Naidu2022}.\footnote{Indications of the  $z>10$ population existed before {\it JWST}. The {\it HST} discovery of the bright (H=26.0) galaxy GNz11 \citep{Oesch2016} provided an early hint that extremely luminous galaxies would be present at $z\gtrsim 10$ in relatively small ($\sim 100-300$ arcmin$^2$) survey areas. The 2012 observations of the UDF revealed a faint candidate (H=29.3) with a photometric redshift of $z\simeq 12$ over a smaller survey area, which hinted at a  more gradual evolution in the SFR density at $z>10$ than implied by constant SFE models \citep{Ellis2013}. This galaxy has now been spectroscopically confirmed at the originally-claimed photometric redshift \citep{CurtisLake2023}. These indications were at the limit of what was capable with {\it HST}. } 
The small areas over which these galaxies were identified suggested a much higher number density of luminous $z\gtrsim 10$ galaxies than predicted by many earlier models. As larger areas were obtained, this early result has been confirmed with improved statistics.  And many of the $z\gtrsim 10$ photometric candidates have been confirmed spectroscopically with the low resolution prism mode on NIRSpec (Fig. 2, \citealt[e.g.,][]{CurtisLake2023, Bunker2023,ArrabalHaro2023, Fujimoto2023, Harikane2024, Carniani2024}). Current measurements of the $z\gtrsim 10$ luminosity function now extend over a large dynamic range in absolute magnitude, ranging from M$_{\rm{UV}}=-22$ to $-17$ (Fig. 3a). The $z\gtrsim 9$ photometric samples are large ($\gtrsim 200$ across $\sim 200-400$ arcmin$^2$), with many extremely faint  (m$_{AB}$=29-30) galaxies (e.g., \citealt{Donnan2024}, Whitler et al. 2024b). These measurements reveal that the number density of galaxies (and $\rho_{\rm{UV}}$) slowly declines between $z\simeq 9$ and $z\simeq 12.5$ (Fig. 3b). 
The redshift evolution in the SFR density is more gradual than predicted by the constant SFE models. \citet{Donnan2023} derived a log-linear relation, $\log_{10} (\rho_{UV}) = -0.231+0.037z+(27.5\pm0.3)$, consistent with most subsequent {\it JWST} investigations \citep[e.g.,][]{PerezGonzalez2023, Harikane2023, Adams2024, Finkelstein2024, Mcleod2024, Donnan2024} . The integrated UV density in galaxies with M$_{\rm{UV}}<-17$ at $z\simeq 12.5$ is roughly 7$\times$ larger at $z\simeq 12.5$ than expected prior to {\it JWST} based on the constant SFE extrapolations. 
The latest work has extended these efforts to $z\simeq 14$, with spectroscopy recently confirming the presence of  galaxies at $z\simeq 13.9$ and $z\simeq 14.2$ \citep{Carniani2024}. The first measurements of the $z\gtrsim 12$ luminosity function indicate a continued slower-than-expected decline in the number density of galaxies \citep[e.g.,][]{Donnan2024,Robertson2024}. 
Above $z\simeq 15$, there have yet to be any galaxies   confirmed spectroscopically. 

The $z\gtrsim 10$ census from {\it JWST} has revealed more galaxies than predicted by many 
theoretical models. What is causing the excess UV emission is not yet clear. Several early papers 
proposed that the SFE may be larger at $z\gtrsim 10$, perhaps reflecting suppressed feedback in early 
galaxies \citep{Dekel2023, Qin2023, Wang2024}.  The effects of top-heavy IMFs on the L$_{\rm{UV}}$/SFR ratios have also been investigated in several studies \citep{Trinca2024, Ventura2024, Hutter2024}. \citet{Ferrara2023} suggested that the slow evolution at the bright end of the luminosity function could be explained by a decrease in  attenuation at earlier times, driven by the ejection of  dust via radiation pressure  \citep{Ziparo2023}. Others demonstrated that the $z\simeq 10-12$ UVLF can be reproduced if there is a large dispersion ($\sigma_{\rm{MUV}}$) in the M$_{\rm{UV}}$-M$_{\rm{halo}}$ relation, allowing low mass galaxies to upscatter to large luminosities (e.g., \citealt{Mason2023,Shen2023,Sun2023,Mirocha2023}). Sources of UV scatter include halo assembly, temporal fluctuations in star formation histories (SFHs), and variations in dust or ionizing spectrum (L$_{\rm{UV}}$/SFR). The scatter due to halo assembly reflects variations in halo growth rates at fixed mass, resulting in an estimated 1$\sigma$ scatter of $\sigma_{\rm{halo}}=0.3$ dex \citep[e.g.,][]{Ren2019, Mirocha2023}. A larger effect may be expected from variable (or `bursty') SFHs, which are thought to arise naturally at high redshift  (\S\ref{sec:sfhs}, \citealt[][]{FaucherGiguere2018}). 
Most studies show that a moderate value of scatter ($\sigma_{\rm{MUV}}\simeq 0.75$) can reproduce the results at $z\lesssim 12$. This may help explain why high-resolution galaxy formation simulations (which often include significant UV variability) are often better able to reproduce the UVLF at $z\simeq 10-12$ than  simple models with no scatter \citep[e.g.,][]{Keller2023, Sun2023, Kannan2023, McCaffrey2023}. Most recently it has been demonstrated that even larger UV variability ($\sigma_{\rm{MUV}}\simeq 1-2$) may be required to explain the UVLF at $z\simeq 12-14$
\citep{Shen2023, Sun2023, Gelli2024,Kravtsov2024}. It is not yet clear whether such a sudden shift is consistent with observations. \citet{Feldmann2024} argue that the UVLFs at $z\simeq 6-14$ are naturally reproduced in the context of the (non-evolving) SFE-halo mass relation predicted in the FIREBox high resolution simulations. To break these degenerate explanations for the $z>10$ UVLF, it will be necessary to move beyond the UVLF. Measurements of UV variability in individual galaxies (\S\ref{sec:sfhs}), galaxy clustering (\S\ref{sec:clustering}), and reionization (\S\ref{sec:reionization}) offer complementary constraints on the relative importance of UV scatter and SFE in driving the early UVLF \citep{Munoz2023,Nikolic2024,Gelli2024}.

One of the remaining frontiers is at the faint end of the luminosity function. Both {\it HST} and {\it JWST} only probe down to M$_{\rm{UV}}\simeq -17$ in conventional deep field observations. The flattening  in the luminosity function is expected from a variety of  processes  that suppress star formation in the lowest mass galaxies \citep[e.g.,][]{Kuhlen2012, Jaacks2012, Wise2014, Wu2024}. The absolute magnitude at which this flattening or turnover appears is not well known. Theoretical predictions  
suggest that the turnover may arise somewhere between M$_{\rm{UV}}\simeq -12$ and $-15$. Observational knowledge of the faint galaxy turnover is critical not only to  understanding 
the physics that regulates star formation in low mass halos, but also for predicting the contribution of galaxies to reionization. Gravitational lensing provides the only direct path to constraining the number density of high redshift galaxies in this absolute magnitude regime. The magnification provided by strong lensing can often reach in excess of a factor of 10, allowing ultra-deep imaging to reach more than 2.5 mags below the sensitivity limits in conventional deep fields. The {\it HST} Frontier Fields (HFF) program \citep{Lotz2017} was designed to achieve this goal at $z\simeq 6-8$, investing 840 orbits into deep optical and near-infrared imaging (m$\simeq 29$) of six galaxy clusters. These efforts allowed the $z\simeq 6$ luminosity function to be extended to at least M$_{\rm{UV}}\simeq -15$ \citep[e.g.,][]{Livermore2017, Atek2018, Bhatawdekar2019, Bouwens2022lf}.  The faint end slope was found to be steep ($\alpha\simeq -2$), similar to that found in non-lensing fields. The data are consistent with a turnover at  M$_{\rm{UV}}\sim -15$, although systematic uncertainties are large below this limit \citep{Bouwens2017, Atek2018}.  {\it JWST} has recently conducted an analogous program, obtaining ultra-deep imaging of Abell S1063 (also known as RXCJ2248). Analysis of these data will soon offer new insight into the turnover in the UVLF at $z>9$.

\begin{figure}[t]
\centering
\includegraphics[width=0.95\textwidth]{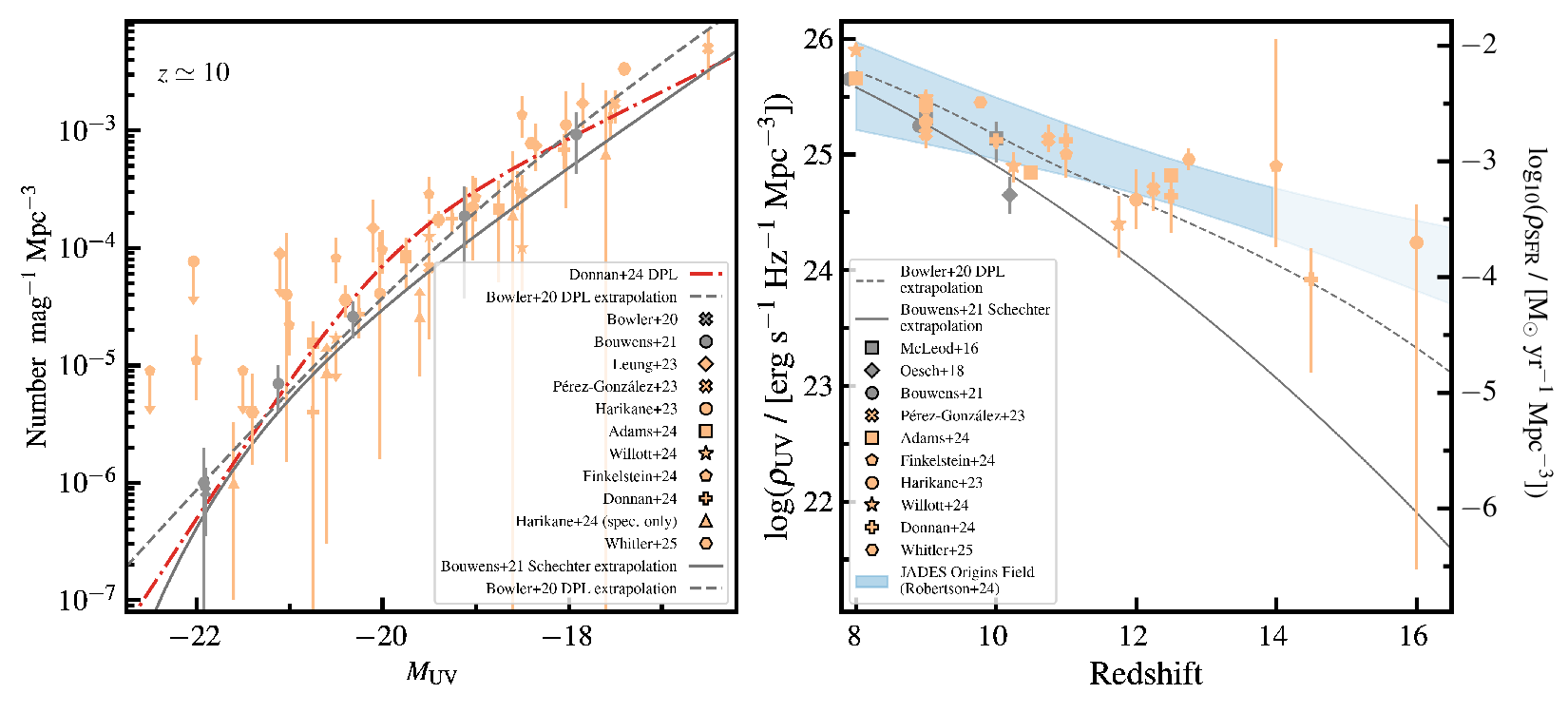}
\caption{(Left:) UV luminosity function at $z\simeq 10$, with large number of constraints from different analyses of NIRCam imaging in many different fields. (Right:) UV luminosity density (M$_{\rm{UV}}<-17)$ at $z\simeq 8-16$, demonstrating an excess above pre-{\it JWST} extrapolations. Both plots are adapted from \citet{Whitler2025}, with Schechter and DPL parametric LFs as derived in that work.}
\label{fig:uvlf}
\end{figure}

\begin{table}[t]
\TBL{\caption{Luminosity function parameters for double power law function at $z=9-15$  from \citet{Donnan2024}. Integrated UV luminosity density is reported in rightmost column with an integration limit of M$_{\rm{UV}}=-17$. }\label{tab:uvlf}}
{\begin{tabular*}{\textwidth}{@{\extracolsep{\fill}}@{}lccccc@{}c}
\toprule
\multicolumn{1}{@{}l}{\TCH{$z$}} &
\multicolumn{1}{c}{\TCH{$\rm \phi^*/mag/Mpc^2$}} &
\multicolumn{1}{c}{\TCH{$\rm M_{AB}^*$}} &
\multicolumn{1}{c}{\TCH{$\alpha$}} &
\multicolumn{1}{c}{\TCH{$\beta$}} &
\multicolumn{1}{c}{\TCH{$\log(\rho_{\rm UV}/\rm erg~s^{-1}Hz^{-1}Mpc^{-3})$}} &
\\
\colrule
$9$    &  $(23.5\pm39.2)\times10^{-5}$ & $-19.70\pm0.96$ & $-2.00\pm0.47$ & $-3.81\pm0.49$ & $25.29^{+0.05}_{-0.05}$\\
$10$   &  $(14.5\pm16.4)\times10^{-5}$ & $-19.98\pm0.61$ & $-1.98\pm0.40$ & $-4.05$        & $25.12^{+0.07}_{-0.08}$\\
$11$   &  $(3.27\pm9.86)\times10^{-5}$ & $-20.73\pm1.61$ & $-2.19\pm0.69$ & $-4.29\pm1.30$ & $25.12^{+0.14}_{-0.20}$\\
$12.5$ &  $(0.99\pm0.99)\times10^{-5}$ & $-20.82\pm0.71$ & $-2.19$        & $-4.29$        & $24.64^{+0.18}_{-0.32}$\\
$14.5$ &  $(0.18\pm0.28)\times10^{-5}$ & $-20.82$        & $-2.19$        & $-4.29$        & $23.92^{+0.27}_{-0.81}$\\
\botrule
\end{tabular*}}{}
\end{table}
\subsection{The Contribution of Dusty Star Forming Galaxies at $z>4$}\label{sec:dustSFRD}

For more than a decade, substantial effort has been invested in determining the level of obscured star formation activity at $z\gtrsim6$ not directly captured by the UVLF (see \citealt{Casey2014} for a review).
In the run up to {\it JWST}, a collection of blind 2--3 mm ALMA surveys implied that $\sim$30\% of star formation was obscured at $z\sim5-6$, with the obscured fraction declining with increasing redshift \citep{Zavala2021}.
Complementary efforts utilized targeted ALMA follow-up of UV-luminous ($-23 \lesssim M_\mathrm{UV} \lesssim -20$) $z\sim4-7$ galaxies to estimate the level of obscured star formation in relatively common high-redshift galaxies.
From various analysis methods, these studies determined that the total fraction of obscured star formation was around $\sim$60\% at $z\sim5.5$ \citep{Khusanova2021} and $\sim$30\% at $z\sim7$ \citep{Algera2023,Barrufet2023_REBELS}.
Moreover, \citet{Algera2023} concluded that approximately half of star formation within massive ($M_\ast\,\sim10^{10}\,M_\odot$) $z\sim7$ galaxies is obscured on average, much less than that among similarly-massive galaxies at $z\lesssim2.5$ \citep{Whitaker2017}.
Overall, while these $z\gtrsim6$ obscured fraction estimates had large uncertainties, the overall consensus prior to {\it JWST} was that a significant fraction of cosmic star formation remained obscured to at least $z\sim7$, though the majority of star formation in the first Gyr was likely unobscured by dust. 

Shortly after the first {\it JWST}/NIRCam data were delivered, several faint (F150W$\gtrsim$27 AB) objects were identified with very red colors between observed-frame 1.5$\mu$m and $\approx$4$\mu$m \citep[e.g.,][]{Barrufet2023,PerezGonzalez2023,Rodighiero2023}.
The majority ($\sim$70\%) of these so-called {\it HST}-dark galaxies initially seemed to be dusty star-forming systems at $z\sim2-8$, while the remaining subset appeared to be either quiescent $3<z<5$ galaxies or $z>6$ galaxies with extremely strong nebular line emission contaminating the reddest NIRCam bands \citep{PerezGonzalez2023}.
Early work concluded that the dusty star-forming subset contributed an approximately constant obscured star formation rate density of $\sim1.5\times10^{-3}$ $M_\odot$/yr/Mpc$^3$ over $3<z<8$, overtaking the aggregate cosmic stellar mass assembly of classical sub-mm galaxies at $z>6$ \citep{Barrufet2023}.
However, as deep overlapping MIRI imaging began to accumulate, the inclusion of this data systematically lowered the SFR and stellar mass estimates of {\it HST}-dark galaxies by $\approx$0.6 dex (up to $\sim$1 dex) compared to early analyses with {\it HST}+NIRCam data alone \citep{Williams2024}.
Moreover, as spectroscopic follow-up began to accumulate, it became clear that the majority of {\it HST}-dark galaxies at $z>6$ are a population of broad-line AGN that remain poorly understood (\citealt{Greene2024,Matthee2024,Williams2024}; see \S\ref{sec:AGN}).
If the very red optical continua of these peculiar $z>6$ broad-line AGN are primarily due to attenuated emission from black hole accretion rather than obscured star formation, estimates of the obscured star formation rate density at $z>6$ could fall dramatically \citep{Williams2024}.
Dedicated ALMA+{\it JWST} follow-up of these systems is currently underway to better assess their obscured and unobscured star formation activity.

Early studies with {\it JWST} data have also begun to assess the extent to which red galaxies contribute to the cosmic stellar mass density at $z>4$.
Initial results suggest that red galaxies (defined as those with rest-frame UV continuum slopes of $\beta > -1.2$; see \S\ref{sec:uvslopes}) comprise $\sim$50\% of galaxies with stellar mass $M_\ast = 10^{10}\,M_\odot$ at $z=4-8$ \citep{Weibel2024_SMF}.
Moreover, it appears that {\it HST}-based Lyman-break selections missed $\sim$20\% of these relatively massive galaxies at $z\sim4$ and as many as 40--50\% at $z\sim6-8$ \citep{Gottumukkala2024,Weibel2024_SMF}. 
While these red galaxies dominate the high-mass end ($M_\ast \gtrsim 10^{10}\,M_\odot$) of the stellar mass function at $z\sim4-5$, bright blue objects traditionally selected in the UV seem to begin dominating at $z>6$, at least assuming that the red optical continua of the peculiar broad-line AGN mentioned above are due to attenuated emission from black hole accretion rather than stellar emission \citep{Weibel2024_SMF}.

\section{Star Formation Rates, Stellar Masses, and Stellar Population Ages}\label{sec:galaxyprops}

Prior to {\it JWST}, our understanding of early galaxy properties was mostly derived from broadband SEDs combining  photometry from  {\it HST} and the {\it Spitzer Space Telescope}.  At $z\gtrsim 5$, {\it HST} filters are limited to the rest-UV, with the light output largely powered by recently formed O and B stars. The far-UV continuum provides information on the recent star formation rate (if coupled with a correction for dust attenuation), but it does not constrain the timescale over which star formation has been ongoing  or provide a measure of the total stellar mass (M$_{\rm{\star}}$). 
The Infrared Array Camera (IRAC) on  {\it Spitzer} provided our first view of the rest-frame optical at $z\gtrsim 5$ with 
its two most sensitive broadband filters centered at 3.6$\mu$m and 4.5$\mu$m. It is at these redder wavelengths where the presence of older A stars can be seen  in early galaxies, constraining the light-weighted age of the stellar population as well as the total stellar mass.

In practice, stellar masses and SFRs are typically inferred by comparison  of observed galaxy SEDs to stellar population synthesis models (see \citealt{Conroy2013} for a review). 
In what follows, we will describe the  light-weighted stellar population ages of $z\gtrsim 5$ galaxies, and here we briefly describe the observational signatures in UV and optical photometry. We will refer to galaxies as having ``old'' ages if their SEDs have  morphologies consistent with  $\simeq 100$ Myr of constant star formation. At these older ages,  a Balmer break will be present, with the integrated light from A stars in the rest-frame optical more luminous than  the rest-UV continuum powered by O and B stars. Young stellar populations (which we define as those formed from $\lesssim 10$ Myr of constant star formation) will be expected in galaxies undergoing a phase of rapid stellar mass growth, where specific star formation rates  (sSFR = SFR/M$_{\rm{\star}}$) are large. Such galaxies will not have built up the luminous A star population required to produce a  Balmer/4000~\AA\ break, and will instead have significant contributions from nebular continuum emission throughout the rest-UV to optical.  Their SEDS will  be characterized by  a Balmer jump at 3645~\AA\ (from free-bound nebular continuum) and extremely strong emission lines in the rest-frame optical ([OIII], H$\beta$, H$\alpha$). Since the rest-frame optical continuum remains very weak at young ages, the emission lines will have  large equivalent widths (EWs), often dominating the light in  broadband (and medium band) filters. The measured flux density in filters that are contaminated by emission lines will be brighter than neighboring filters that are only sensitive to the weaker continuum emission. These emission line ``flux excesses'' are one of the key signposts of galaxies with young stellar populations and large sSFR. 

In the era of {\it HST} and {\it Spitzer}, most work on SEDs was focused on bright galaxies (M$_{\rm{UV}}<-20$) owing to the limited sensitivity of the deepest IRAC images (typically m$_{\rm{AB}}\sim26.0$ at 5$\sigma$). While some Balmer Breaks were thought to be present with {\it Spitzer} \citep[e.g.,][]{RobertsBorsani2020,Whitler2023a}, most SEDs were shown to have the characteristic flux excesses associated with strong rest-optical emission lines \citep{Stark2013, deBarros2014, Labbe2013, Smit2018}. This suggested that bright $z\gtrsim 6$ galaxies tend to be in a phase of rapid assembly with large sSFR ($\gtrsim 10$ Gyr$^{-1}$) and SEDs dominated by young stellar populations. 
Based on {\it Spitzer}/IRAC flux excesses, the [OIII]+H$\beta$ EW distribution was derived at $z\simeq 7$ (Fig. 4a), revealing a median value of 760~\AA{} \citep{Endsley2021}, consistent with a light-weighted stellar population age of 75 Myr (assuming constant star formation history, hereafter CSFH).\footnote{For simplicity, we will refer to light-weighted stellar population ages  derived assuming constant star formation histories. These are meant to provide benchmark values for relative comparison, but we note these are unlikely to correspond to absolute measures of the age of the stellar population. As explained in this section, the true stellar population age will vary under different assumed star formation histories.  } Twenty percent of  UV-bright $z\simeq 7$ galaxies were shown to have younger SEDs ($\lesssim$ 30 Myr) with [OIII]+H$\beta$ EWs in excess of 1200~\AA\ and sSFRs greater than 30 Gyr$^{-1}$, indicating a recent  upturn or burst of star formation \citep{Endsley2021}. These results gave our first observational indication that star formation was likely bursty at $z\gtrsim 6$, similar to  predictions of many galaxy formation simulations (discussed in \S\ref{sec:sfhs}). Bursty star formation histories also  have significant consequences for our ability to reliably measure stellar masses. As galaxies undergo strong bursts, they form bright stellar populations which can outshine fainter older stars that  dominate the mass of the galaxy 
\citep[e.g.,][]{Leja2019}.  If the population of older stars is not accounted for, the stellar mass will be  underestimated. 
For galaxies with SEDs dominated by extremely young stellar populations ($<$10 Myr), it has been shown that the stellar masses carry order of magnitude uncertainties \citep[e.g.,][]{Tacchella2022,Whitler2023a,Narayanan2024a}, with the CSFH models providing a lower bound on the total stellar mass. Non-parametric star formation histories allow the potential presence of older populations to be included in SED modeling, but in most cases the data do not  
constrain  whether or not an old population is present, and  the recovered stellar masses still depend sensitively on the chosen star formation history prior.  As more $z\simeq 5-6$ galaxies are analyzed jointly with NIRCam and MIRI, the potential presence of older stars  should be better constrained \citep[e.g.,][]{Papovich2023}. 

In the last two years, {\it JWST} has made a major improvements on the earlier investigations with {\it HST} and {\it Spitzer}. The sensitivity of the deepest NIRCam images at 2-5$\mu$m (m$_{\rm{AB}}$=31 at 5$\sigma$) enable characterization of galaxies more than 50-100$\times$ fainter than was possible with {\it Spitzer}. The wavelength coverage of NIRCam constrains the rest-frame optical up to $z\simeq 9$, so the most robust work on stellar masses, ages, and SFRs requires focus on the $z\simeq 6-9$ galaxy population. At yet higher redshifts, MIRI (with its longer wavelength coverage) is beginning to contribute meaningfully \citep[e.g.,][]{Hsiao2024,Zavala2024,AlvarezMarquez2024}, but for most $z\gtrsim 9$ systems, the rest-optical SEDs and stellar masses are not well-constrained. We thus will focus our discussion on what has been learned at $z\simeq 6-9$. Galaxy SEDs at these redshifts have now been analyzed by numerous teams in several different deep fields with NIRCam coverage \citep[e.g.,][]{Endsley2023, Endsley2024, Rinaldi2023, Dressler2023, Trussler2024, Harshan2024, Asada2024, Ciesla2024, Clarke2024}. As noted above, the inferred stellar mass depends on the assumed star formation history. For CSFHs, typical stellar masses are found to vary from  M$_{\rm{\star}}\sim10^{9.5}$ M$_\odot$ at the bright end of the luminosity function (M$_{\rm{UV}}=-22$) to M$_{\rm{\star}}\sim10^{7.0}$  M$_\odot$ at the faint end (M$_{\rm{UV}}=-17$). The average SED-derived SFRs (again assuming CSFH) vary from SFR$\sim$50 M$_\odot$ yr$^{-1}$ (M$_{\rm{UV}}=-22$) to SFR$\sim$0.2 M$_\odot$ yr$^{-1}$ (M$_{\rm{UV}}=-17$) (see Fig. 4b).
The [OIII]+H$\beta$ EW distribution has been computed from the NIRCam flux excesses in large samples of $z\simeq 6-9$ galaxies.
For luminous systems, the results are similar to what had been seen prior to {\it JWST}. At  M$_{\rm{UV}}=-20$, the median [OIII]+H$\beta$ EW is large (780~\AA). Similar inferences can be made from the NIRCam grism, where large samples of strong [OIII] emitters have been identified \citep[e.g.,][]{Matthee2023, Meyer2024}. The large EWs suggest that young stellar populations (10-50 Myr) dominate the light of many early galaxies, as found previously with {\it Spitzer}.

Among the less luminous galaxies, the NIRCam flux excesses from [OIII]+H$\beta$ weaken \citep{Begley2024,Endsley2024}. At M$_{\rm{UV}}$=-17.5, the median [OIII]+H$\beta$ EW is 330~\AA, a factor of $\simeq2.5\times$ lower than the values seen at the bright end of the luminosity function. While there are objects with large [OIII]+H$\beta$ EWs among the UV-faint population \citep[e.g.,][]{Atek2024,Endsley2024b}, the overall population appears to exhibit weaker lines. This is likely   partially driven by lower metallicities among the faint galaxies, decreasing the strength of [OIII] emission. Additionally, it has been suggested that star formation histories may also be contributing \citep{Endsley2024}. A subset of UV-faint galaxies are argued to have  experienced a downturn in star formation in the previous 10 Myr, diminishing the strength of rest-optical emission lines. This subset of weak emission line galaxies may correspond to the `off-mode' phase  that follows the bursts of star formation that had been observed years ago by {\it Spitzer}.
The [OIII]+H$\beta$ EW trend with M$_{\rm{UV}}$ suggests that the distribution of star formation histories may be different at the bright and faint end of the luminosity function. At $z\gtrsim 6$, the most UV luminous galaxies are often found with SEDs indicative of a recent upturn in star formation. Among fainter galaxies, there is a wider range of star formation histories, with  
significant fraction having SEDs indicative of a recent a decline in SFR. This picture is consistent with expectations of bursty star formation histories, where galaxies take on different absolute UV magnitudes as they fluctuate in SFR \citep{Endsley2024b, Dressler2024}. We will discuss what we are learning about the specifics of the bursty star formation histories in \S\ref{sec:sfhs}.

{\it JWST} is  facilitating much-improved 
measurement of the relationship between the galaxy SFR and stellar mass (the so-called star forming main sequence, SFMS). It is expected that the normalization of the SFMS (the  specific star formation rate) will increase toward higher redshifts, tracking the larger baryon accretion rates onto dark matter halos \citep[e.g.,][]{Weinmann2011}. 
At $z\simeq 2-3$, the typical specific star formation rates for 10$^{9}$  M$_\odot$ galaxies are 2-3 Gyr$^{-1}$ \citep{Reddy2009}.  At $z\gtrsim 6$, the specific star formation rates have been measured with a variety of SFR indicators. 
Measurements in which the SFR is computed by combination of  unobscured and obscured star formation (using dust continuum detections for the latter) suggest that the median sSFR at $z\simeq 7$ is 17 Gyr$^{-1}$ \citep{Topping2022}. At fixed stellar mass, these authors find that the sSFR  increases at a rate of (1+z)$^{1.7}$ over $1<z<7$, broadly consistent with the rate expected owing to evolution in baryon accretion rates.  {\it JWST} results suggest that fainter $z\gtrsim 6$ galaxies often have even larger sSFR, with typical values of 100 Gyr $^{-1}$ in $z\simeq 6.5-8$ galaxies with M$_{\rm{UV}}=-20$ (assuming CSFH), consistent with a population of galaxies with in the midsts of very strong bursts of star formation.  Care must be taken when interpreting these large sSFR values, as they are subject to the outshining problem described above. Different assumptions on the SFH  can fit the data with larger stellar masses, reducing the sSFR by factors of several. Measurement of the SFMS faces additional challenges from the difficulty in constructing mass-complete samples. The lowest mass galaxies in magnitude-limited samples will always be biased to young systems (owing to their larger light-to-mass ratios). For example, an M$_{\rm{\star}}=10^{7}$ M$_\odot$ $z\simeq 7$ galaxy will be fairly bright if it is dominated by young stars (m$_{\rm{AB}}=28.4$ in rest-optical at 3 Myr), but will be too faint to be observed by most surveys if it is somewhat older (m$_{\rm{AB}}=30.8$ in rest-optical at 100 Myr). This problem will be further exacerbated if galaxies have undergone recent downturns in star formation \citep{Endsley2024b}, causing  systems to appear extremely faint after $\simeq 30$ Myr of low SFR activity. As a result, continuum-selected samples will often not sample the full range of SFHs among low mass galaxies \citep{Sun2023completeness}. This bias can artificially boost the average sSFR at low masses if  not taken into account. 

The extremes of the stellar mass distribution represent frontiers of considerable interest in the {\it JWST} era.  In cluster fields magnified by gravitational lensing,  extremely faint (M$_{\rm{UV}}=-12$) $z\gtrsim 6$ sources are now being characterized with stellar masses as low as $\sim 10^4$ M$_\odot$, pushing well into regime occupied by star cluster complexes \citep{Vanzella2024}. We will discuss these results in more detail in \S\ref{sec:sizes}. At the other extreme, efforts are focused on identifying the emergence of the first massive galaxies (M$_{\rm{\star}}\gtrsim 10^{10}$ M$_\odot$)  appear. The first NIRCam images revealed a set of $z\gtrsim 7$ galaxies with very red rest-optical continuum SEDs, leading to suggestions that there may be a population of extremely massive galaxies  at $z\gtrsim 6$ \citep{Labbe2023}.  While  several of these galaxies now appear to be powered by AGN in the rest-optical (and not old stars)  \citep{Kocevski2023}, some still potentially show Balmer breaks in their spectra (\citealt{Wang2024_RUBIES}; c.f., \citealt{Inayoshi2024}). More investigation is needed to characterize these sources in the future. Wider area surveys will soon enable the high-mass end of the $z\gtrsim 7$ galaxy distribution to be studied in greater detail \citep[e.g.,][]{Casey2023}.

\begin{figure}[t]
\centering
\includegraphics[width=.9\textwidth]{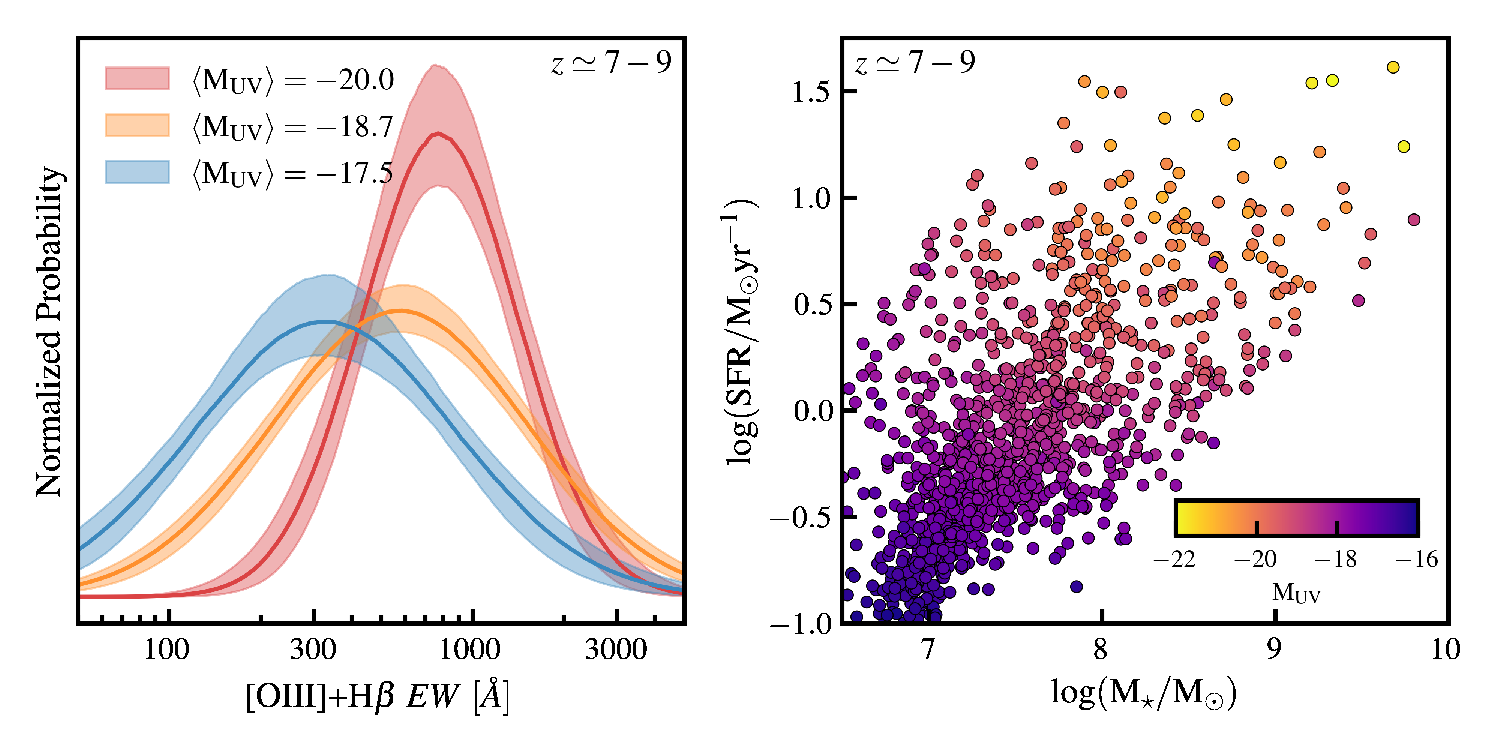}
\caption{(a) [OIII]+H$\beta$ equivalent width distribution in galaxies at $z\simeq 6-9$. The most luminous galaxies power the largest [OIII]+H$\beta$ EW emission lines. Extremely faint galaxies have a broader range of [OIII]+H$\beta$ EWs, likely reflecting lower metallicities and a more diverse range of star formation histories. (b) Star-forming main sequence derived from NIRCam SEDs of $z\simeq 7-9$ galaxies. Both plots are adapted from  \citet{Endsley2024}. }
\label{fig:ewdist}
\end{figure}

\section{Star Formation Histories}\label{sec:sfhs}

In the last ten years, high resolution galaxy formation simulations have led to  revised predictions of how star formation proceeds in high redshift galaxies. It has been found that when the  resolution of the simulation allows gravitationally bound clouds and stellar feedback to be resolved (spatially and temporally), star formation histories of high redshift galaxies are often  highly time variable, with strong bursts of star formation punctuated by quiescent periods (or lulls) of much weaker star formation activity.  An analytic model for the origin of bursty star formation is presented in \citet{FaucherGiguere2018}, and the reader is directed here for more detailed discussion. Strong bursts are expected in cases where the dynamical time is too short ($\lesssim 20$ Myr) for stellar feedback to respond effectively to gravitational collapse.  This is further amplified when star formation is  limited to a small  number of gravitationally-bound clouds, naturally leading to discrete star formation peaks as the individual clouds undergo collapse. \citet{FaucherGiguere2018} demonstrate that both effects  occur in high redshift galaxies, with high gas densities leading to short freefall times and high gas fractions limiting star formation to a small number of clouds. As a result, they argue that significant SFR variability is expected in galaxies at $z\gtrsim 1$.

Observationally star formation histories can be probed by SFR indicators that are sensitive to different timescales. The UV stellar continuum is sensitive to O and B stars, probing star formation over up to $\sim 100$ Myr timescales. Emission lines (and nebular continuum) are powered by O stars, providing an indication of the recent ($\lesssim 10$ Myr) star formation. Measurement of the scatter in the SFMS at $z\simeq 1-3$ has long indicated that star formation histories are likely fairly smooth when the SFR is averaged on 100 Myr timescales \citep[e.g.,][]{Noeske2007, Reddy2012, Rodighiero2014}. In relatively massive $z\simeq 2$ galaxies, no strong evidence for SFR variability was found in the first statistical H$\alpha$ samples, with 
uncertainties in attenuation laws making robust conclusions challenging 
\citep{Shivaei2015}.  But as noted in \S\ref{sec:galaxyprops}, hints of strong bursts became apparent in lower mass $z\gtrsim 4$ galaxies as {\it Spitzer}/IRAC began providing a window on rest-optical emission lines. The extremely large H$\alpha$ or [OIII]+H$\beta$ EWs found in these early investigations 
are very challenging to explain without a recent strong upturn in star formation. To fully validate the bursty star formation picture, it is necessary to also identify the  subset of galaxies that are in-between bursts, with weaker emission lines. This population is likely to also be very faint in the continuum: if star formation shuts off entirely following a burst, a galaxy will become 3 mags fainter in M$_{\rm{UV}}$ in just 30 Myr \citep{Endsley2024b}. Hence a galaxy that was m$_{\rm{AB}}=27$ during its burst will appear as m$_{\rm{AB}}=30$ in a quiescent phase 30 Myr after a burst. This drop in continuum luminosity density can be mitigated somewhat if there is residual SFR between bursts, but   galaxies in lulls of SFR are nevertheless likely to be more common at the faint end of existing surveys.   Robust identification of weak emission lines in faint early galaxies was not possible prior to {\it JWST} given the flux limits of {\it Spitzer}/IRAC photometry. 

{\it JWST} has been a game-changer for characterization of rest-optical lines, leading to a series of detailed investigations of star formation histories at high redshift. After the first deep NIRCam images were released, it was noted that a significant subset of $z\simeq 5-8$ galaxies had SEDs with medium-band flux excesses revealing much weaker [OIII]+H$\beta$ emission lines than had been seen with {\it Spitzer}. These systems appeared to also have SEDs indicating relatively young stellar populations, where the emission lines should have been much stronger. This SED morphology is exactly what would be expected from galaxies that have recently (in last 3-10 Myr) undergone a decline in star formation, with a reduced early O star population leading to the diminution of the emission line strengths.\footnote{Other interpretations for weak emission lines in young galaxies do exist. In particular, if ionizing photons escape galaxies (without being reprocessed into emission lines) the SEDs can be explained without the need for recent SFR declines. The large escape fractions required to explain the weak lines ($>$0.5) coupled with only modest UV slopes $\beta\simeq -2$ in some cases, may indicate that SFR declines are a more natural interpretation. Deep spectroscopy of bright examples should enable these two different pictures to be distinguished, as O star signatures should be readily visible in the escape fraction case.}
Spectroscopy quickly began directly confirming examples of $z\gtrsim 5$ galaxies very weak emission lines, allowing improved constraints on star formation histories  (Fig.~ 5 \citealt{Strait2023, Looser2023, Weibel2024, Endsley2024}). Currently fewer than 10 mini-quenched galaxies have been spectroscopically confirmed. Roughly half correspond to relatively recent ($<$10 Myr) declines in SFR, and the other half often show spectra indicative of a decline in the last 10-30 Myr. Because the weak emission line population appears preferentially at the faint end of the luminosity function, building larger spectroscopic samples will require focus on the intrinsically faintest galaxies. In principle, these investigations offer the potential to constrain the duration of the off-mode phase. If a significant fraction of early galaxies cycle through quiescent periods for 50-100 Myr (and the residual SFR is very low), it should be possible to find examples in upcoming surveys (see \citealt{Weibel2024}).\footnote{If the SFR between strong burst  is non-negligible, it will be more challenging to reliably identify galaxies that have been in a lull for a longer period ($>$50-100 Myr) following a burst.}  As larger samples of extremely faint ($\rm{m_{AB}>29-30}$) galaxies are observed, more mini-quenched galaxies will be uncovered and the duration between bursts will be better constrained observationally.

The detailed SFHs of individual  galaxies described above are now being combined with statistical investigations of  large samples of galaxies with well-characterized SEDs and rest-optical emission lines. Several approaches are possible. The ratio of the H$\alpha$ and the UV continuum   has long been used as a metric of the star formation history, with galaxy samples expected to present large dispersion in their H$\alpha$/UV ratios if strong bursts are present \citep[e.g.,][]{Weisz2012, Emami2019}. The scatter in the SFMS (derived using H$\alpha$ as the SFR indicator) and the H$\alpha$ EW distribution provide analogous measures of burstiness and its dependence on galaxy properties. Current work with all three metrics points to significant SFR scatter in the $z\gtrsim 5$ galaxy populations, supporting the burstiness picture \citep[e.g.,][]{Endsley2024, Endsley2024b, Dressler2024, Ciesla2024, Clarke2024}. The next step is to investigate the  scatter in UV absolute magnitude  at fixed mass, as well as the dependence of the UV scatter on  mass and redshift.
As we discussed in \S\ref{sec:lf}, the UV scatter  needs to be large ($\sigma_{\rm{MUV}}\simeq 0.75$) to reproduce the UVLF at $z\simeq 10-12$, potentially increasing further ($\sigma_{\rm{MUV}}\simeq 1-2$) to explain the UVLF at $z\simeq 12-14$ \citep{Sun2024, Gelli2024, Kravtsov2024}. Observations can directly constrain the UV scatter at fixed stellar mass (not fixed halo mass). Reliable measurements are not trivial however, as 
magnitude-selected samples will be biased against the faint off-mode galaxy population  (see discussion in \S\ref{sec:galaxyprops}). If care is not taken to limit analysis to mass-ranges which are more complete to diverse SFHs (upturns and downturns), the UV scatter is likely to be underestimated.  The first investigations with {\it JWST}  suggest significant UV scatter  ($>$1 mag) is present at fixed stellar mass in high redshift galaxies \citep{Ciesla2024}.  It is additionally possible to use observed metrics of recent SFHs (i.e., H$\alpha$/UV ratios) to constrain the scatter at fixed halo mass, providing a more direct comparison with models. Joint fits of the UVLF and M$_{\rm{UV}}$-dependent stellar population properties will soon enable 
$\sigma_{\rm{MUV}}$ to be constrained as a function of halo mass (specifically constraining the contribution from time variable star formation), providing a test of whether the data are consistent with the large UV scatter required to explain the UVLF.

\begin{figure}[t]
\centering
\includegraphics[width=1\textwidth]{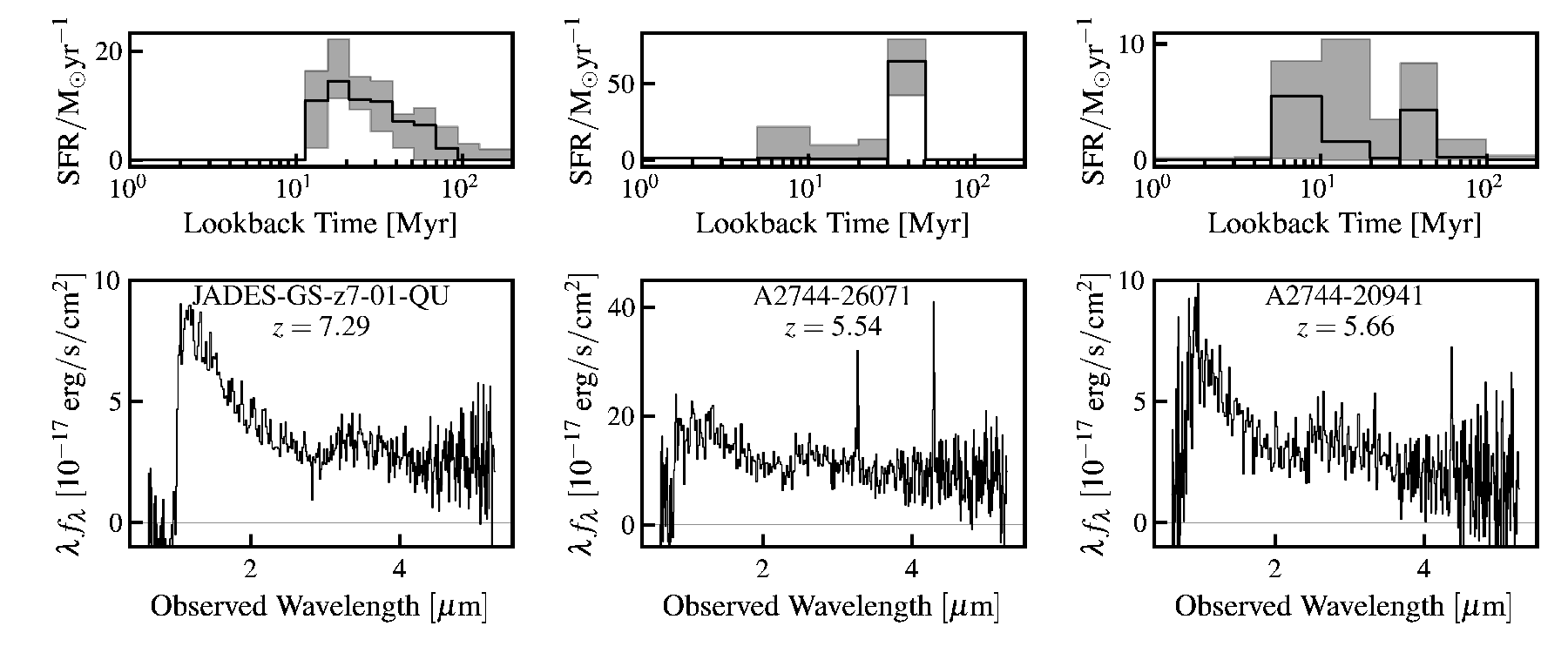}
\caption{Galaxy spectra consistent with lulls in SFR activity (bottom row), as presented in \citet{Endsley2024b}. The combination of weak emission lines and young stellar continuum  are well-fit by a recent (7-10 Myr) downturn in star formation (top row). Such spectra provide compelling evidence for bursty star formation histories in high redshift galaxies.  }
\label{fig:offmode}
\end{figure}

\section{UV Continuum Slopes: Insight into Dust, Stars, and Nebular Continuum}\label{sec:uvslopes}
 
The slope of the UV continuum (defined as $\beta$ where f$_\lambda\simeq \lambda^\beta$) is sensitive to a range of galaxy properties. The total intrinsic continuum is powered by emission from massive stars and nebular continuum. The intrinsic spectrum is  processed by dust as it passes through the ISM, attenuating and reddening the shape of the spectrum. 
At high redshift,  UV slopes are measured by fitting a power law to the  broadband flux densities of filters that sample the rest-frame UV ($\lambda$= 1250-2500~\AA), taking care to avoid those that sample rest-frame wavelengths below Ly$\alpha$.  Direct measurement of $\beta$ from fits to low resolution {\it JWST} prism spectra are also becoming increasingly common in the literature. 
Finally it is also possible to infer UV slopes by fitting the observed SED with a population synthesis model. However this approach may suffer in cases where there are not adequate models to describe the data. 
Interest in measuring UV slopes is driven by their potential for constraining the level of dust attenuation, as required to convert the observed UV continuum to a measurement of the SFR.  If dust is not  present, we should see SEDs with UV slopes approaching the intrinsic value for stellar and nebular (two photon and free-bound) continuum.  The intrinsic slopes depend on the properties of the stellar population (age, metallicity, IMF) and on the relative contribution of nebular continuum. At the young stellar population ages which are common at $z\gtrsim 6$ ($\lesssim 10$ Myr), nebular emission can contribute up to 50\% of the far-UV continuum flux density \citep[e.g.,][]{Topping2022, Katz2024}. In the near-UV (and blue side of the optical), the nebular continuum fraction rises to 60-80\%. In these cases, the continuum (even at $\gtrsim 3000$~\AA) is mostly that of an HII region.
The nebular continuum spectrum acts to redden the very blue stellar continuum, leading to intrinsic slopes of roughly $\beta\simeq -2.5$ (Fig.~ 6a). As we will discuss below, several  effects modulate the nebular  contribution, allowing the bluer continuum slopes to be observed in specific cases.

Observations  with {\it HST} and  {\it JWST} have characterized UV slopes for large samples of $z\simeq 4-8$ galaxies. The UV-continuum selected population tends to have have fairly blue UV slopes, consistent with modest  attenuation toward the dominant UV-emitting components in $z\gtrsim 4$ galaxies. The UV slope has been shown  to correlate with the absolute magnitude of the galaxy (Fig.~ 6b), with the most luminous galaxies having slightly redder slopes than galaxies at the faint end of the luminosity function. For example, at $z\simeq 7$ the {\it JWST}/NIRCam photometry presented in \citet{Topping2024b} suggests a relationship of $\beta$ and M$_{\rm{UV}}$ with the following form
\begin{equation}
\beta = -0.12\times \rm (M_{UV}+19)-2.26.
\end{equation}
Similar results have been presented in a variety of other investigations \citep[e.g.,][]{Cullen2024, Austin2024, Morales2024}. Physically this relationship suggests that more luminous galaxies tend to face more attenuation, likely a manifestation of the underlying mass-metallicity relationship (see \S\ref{sec:metallicity}), whereby the more luminous galaxies tend to be more massive on average, with 
ISM reservoirs that are more metal and dust-rich.  In \S\ref{sec:dustprops}, we will discuss the relationship between the UV slope and the level of obscured star formation in early galaxies. 

The UV slopes evolve with redshift: the $z\simeq 4-8$ galaxy population tends to be considerably bluer than  galaxies at $z\simeq 2$ with similar continuum luminosity.  At an absolute magnitude of M$_{\rm{UV}}=-19.5$, the median UV slope shifts from $\beta = -1.7$ ($z\simeq 2$) to $\beta = -2.1$ ($z\simeq 8$).  Such  redshift evolution suggests we are seeing less attenuation toward the UV-emitting components at higher redshifts, as might be expected if the galaxies at a given M$_{\rm{UV}}$ are more metal poor and less massive at higher redshifts.  {\it JWST} has  extended UV slope measurements to $z\simeq 9-15$, a redshift regime that {\it HST} was ill-suited to probe given its absence of photometric coverage beyond the H-band. Most investigations have found that the average UV slopes approach the intrinsic value expected for combined stellar and nebular emission at $z\gtrsim 12$ (median $\beta$ = -2.6), even at the bright end of the luminosity function (Fig.~6b, \citealt[e.g.,][]{Cullen2024, Topping2024b, Austin2024, Morales2024, Casey2024}).  Typical galaxies at $z\gtrsim 12$ thus appear negligibly attenuated by dust. This may not be surprising if the galaxy population at these redshifts is low in stellar mass, with UV luminosities elevated owing to strong bursts of star formation (see \S\ref{sec:galaxyprops}). Recent studies have proposed that additional physical mechanisms may be at play. It has been suggested that the radiation pressure in early galaxies may be sufficient during strong bursts to eject dust from the UV-emitting regions \citep[e.g.,][]{Ferrara2023, Ziparo2023}. In addition, the dust-to-stellar-mass ratios may be much lower in the majority of the galaxy population at $z\gtrsim 10$, only increasing to typical values in more massive galaxies at $z\simeq 7$ \citep{Narayanan2024}.

Attention is now  beginning to focus on the  extremes of the UV slope distribution. Many groups have presented examples of $z\gtrsim 4$ galaxies with  UV slopes ranging from $\beta \simeq -3.0$ to $-2.8$, bluer than the intrinsic nebular and stellar continuum (Fig.~6a, \citealt[e.g.,][]{Topping2024b, Cullen2024, Austin2024}). 
To explain such blue UV slopes, the contribution from nebular continuum emission must  be reduced by some process, allowing  the bluer stellar continuum to be observed directly. 
This can be achieved if Lyman continuum (LyC) photon escape fractions (f$_{\rm{esc}}$) are large (i.e., $\gtrsim 0.5$), such that a majority of the ionizing radiation is not reprocessed into nebular emission \citep[e.g.,][]{Zackrisson2013}. 
Alternatively if galaxies are observed shortly after a strong burst of star formation, when the O star population has begun to disappear, the nebular emission contribution will be reduced and UV slopes will reach extremely blue values for a brief window. 
Current work conservatively estimates that 4\% of the population has $\beta \lesssim -2.8$ at $z\simeq 6-11$ \citep{Topping2024b}, where this fraction only includes galaxies that pass various S/N tests on the UV slope measurement (and hence is a lower limit). At $z\gtrsim 10$, three of the first galaxies confirmed spectroscopically exhibit UV slopes with $\beta < -2.8$ \citep{CurtisLake2023,Hainline2024,Hsiao2024}, potentially  suggesting that whatever physical process drives this 
phase may become more common at the highest redshifts currently probed by {\it JWST}. We will come back to comment on this extremely blue population when describing the contribution of galaxies to reionization in \S\ref{sec:galaxycontribution}.

There is also  focus on the subset of $z\simeq 4-7$ galaxies with highly-reddened UV continuum slopes ($\beta\simeq -0.5$ to -1.5).
The red slopes may reflect significant attenuation in some cases, providing a valuable signpost of what are presumably the most massive galaxies at early times (see discussion in \S\ref{sec:dustSFRD}).  
Most notably, an extremely red ($\beta = -0.5$)  galaxy was recently spectroscopically-confirmed at $z=6.73$ in GOODS-N \citep{Shapley2024}. While faint in the rest-frame UV, this galaxy is now clearly recognized as one of the most massive and evolved systems known at $z\simeq 7$.  In some cases, star formation histories may also contribute to the  colors,  as galaxies will redden significantly in the 10-100 Myr after a strong burst of star formation, easily reaching values of $\beta \simeq -1$ \citep{Mirocha2023,Narayanan2024}. While the fraction of extremely red galaxies decreases at lower continuum luminosities, NIRCam has revealed that there are extremely reddened galaxies at the faint end of the UV luminosity function.  This red population does appear to mostly disappear at $z\gtrsim 9$ in existing samples. The distribution of UV colors should offer an additional diagnostic of star formation histories, with the extreme blue and red ends  sensitive to the duration galaxies remain in SFR lulls (and the residual SFR in between bursts). 

\begin{figure}[t]
\centering
\includegraphics[width=.95\textwidth]{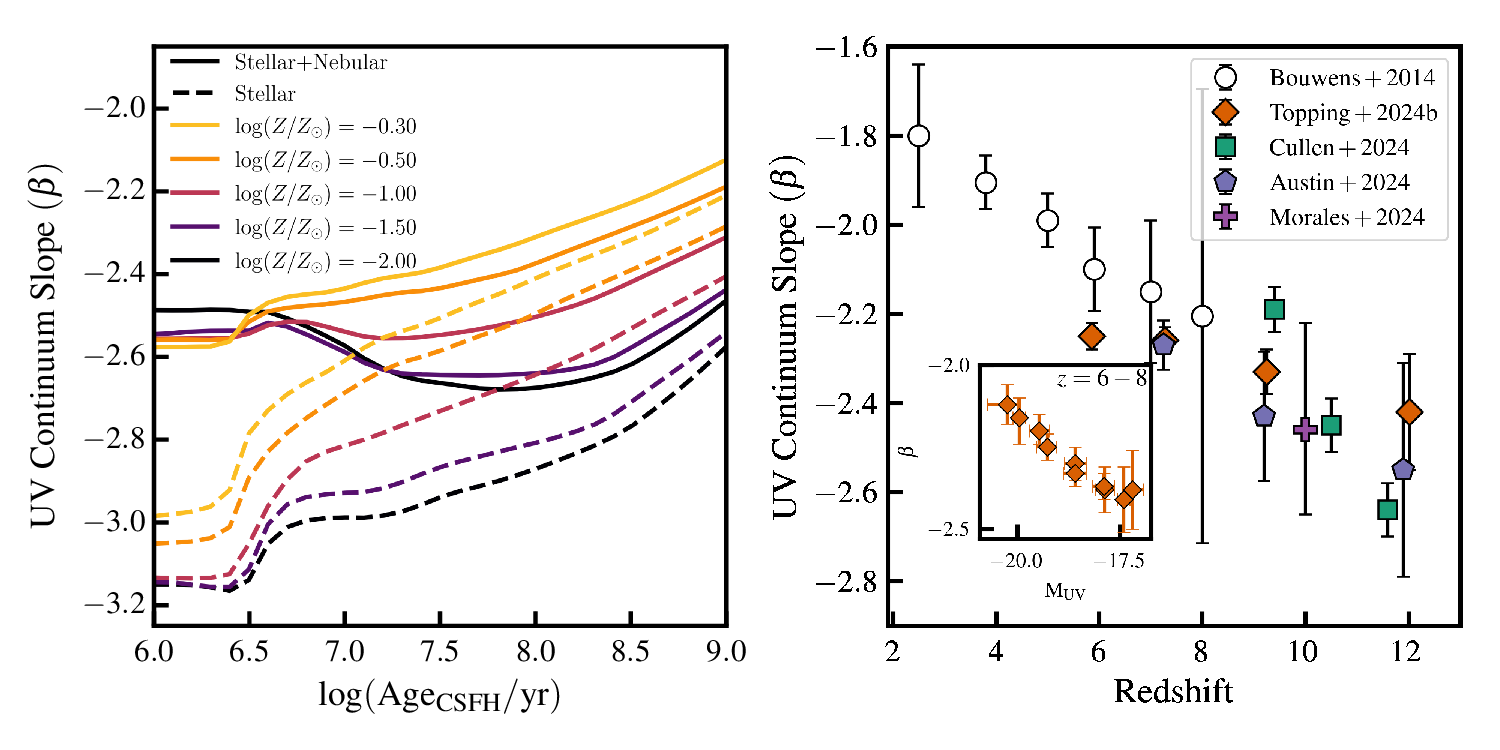}
\caption{(Left:) Intrinsic UV slopes at different stellar population ages and metallicities, adapted from \citet{Topping2022}. Solid lines show models that include contribution from stellar and nebular emission. Dashed lines only include stellar emission. 
(Right:) Observed UV continuum slopes vs redshift from {\it JWST} (colored circles) and {\it HST} (white circles). The observations approach the intrinsic UV slopes of a stellar and nebular emission model at $z\gtrsim 10$, suggesting very little attenuation. The inset shows the relationship between UV slope and absolute magnitude. Ultra faint galaxies also have UV colors that approach the intrinsic $\beta$ value, indicating very little attenuation.  }
\label{fig:beta_evolution}
\end{figure}

\section{Galaxy Sizes, Morphological Structures, and Star Formation Rate Surface Densities}\label{sec:sizes}

Measurements of size and structure provide important insight into how early galaxies are  assembled.  Imaging cameras on {\it HST} delivered the first high spatial resolution view 
of high redshift galaxies. However at the highest redshifts (i.e., $z\gtrsim 7$), useful {\it HST} imaging was only possible in the near-infrared, with WFC3/IR providing 
a typical sampling scale of $\sim 1$ kpc. Without coverage at 2-5$\mu$m, $z\gtrsim 7$ morphological measurements with {\it HST} were only sensitive to the sites of recent star formation probed by the rest-frame UV. 
{\it JWST}/NIRCam imaging delivers  distinct advantages relative to {\it HST}. The spatial resolution is much-improved ($\sim$ 0.1 kpc in the rest-UV at $z\simeq 7$), enabling very high redshift galaxies to be resolved into multiple star forming complexes (or `clumps'). At longer wavelengths (2-5$\mu$m), {\it JWST} has provided the  first high spatial resolution view of the rest-optical in high redshift galaxies, offering the potential to identify the locations of the dominant stellar mass components.

The evolution of galaxy sizes provides one of the most valuable metrics of galaxy assembly. {\it HST} imaging enabled rest-UV sizes to be determined for statistical samples out to $z\simeq 8$. When galaxies were compared at fixed luminosity across a large redshift baseline ($0<z<8$), the half-light radii (R$_e$) were generally found to evolve as R$_e \propto (1+z)^{-\alpha}$ with $\alpha\sim 1$ \citep[e.g.,][]{Ferguson2004, Oesch2010, Shibuya2015}, consistent with expectations if the galaxy size is regulated by the specific angular momentum of the parent  dark matter halo \citep[e.g.,][]{Fall1980, Mo1998}.  This picture is bolstered by abundance matching investigations which show that galaxy sizes are strongly correlated with the halo virial radii (R$_{vir}$), maintaining a nearly fixed ratio  of R$_e$/R$_{vir}=1-3$\% over many orders of magnitude in mass and over a wide redshift range 
\citep[e.g.,][]{Kravtsov2013, Kawamata2015, Shibuya2015, Somerville2018}. These investigations suggest galaxy sizes are at least partially governed by the host halo angular momentum, although  
the complete physical interpretation is likely  more complex  \citep[e.g.,][]{Somerville2018, Jiang2019, Behroozi2022, Karmakar2023}.  

\begin{figure}[t]
\centering
\includegraphics[width=.9\textwidth]{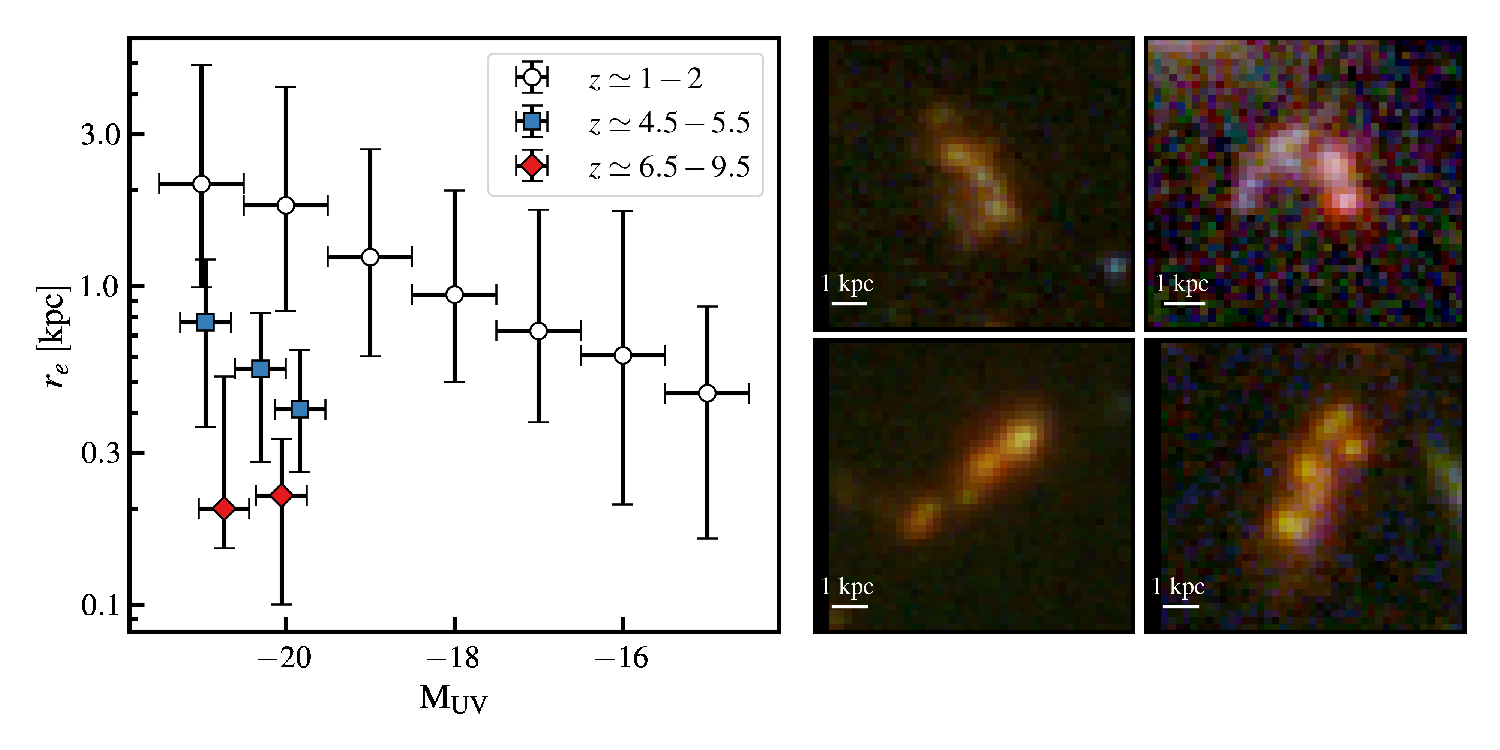}
\caption{(Left:) Evolution in galaxy sizes at fixed absolute UV magnitude. The $z\gtrsim 4.5$ NIRCam measurements are derived in \citet{Ono2024}. For comparison, we show $z\simeq 1-2$ {\it HST} measurements from \citet{Shibuya2015}. The evolution toward smaller sizes at higher redshifts results in very high stellar mass densities within galaxies at $z\gtrsim 5$.  (Right:) NIRCam color images of $z\gtrsim 5$ galaxies, demonstrating the clumpy $\gtrsim 200$ pc substructures that are often present in the brightest galaxies. NIRCam pixel-by-pixel SEDs allow ages and stellar masses to be inferred for individual clumps. Early galaxies appear to mostly be in very early stages of inside-out growth, building up compact bulges.  }
\label{fig:sizes}
\end{figure}

{\it JWST} has extended size evolution measurements to higher redshifts and lower luminosities. The observational results are consistent with what was seen with {\it HST} \citep[e.g.,][]{Yang2022, Sun2024, Morishita2024,Ono2024,Ormerod2024}. Galaxies are very small at $z\gtrsim 5$ (Fig.~\ref{fig:sizes}a), with median half-light radii of $0.1-0.2$ kpc at $z\simeq 6.5-9.5$. A significant fraction of the galaxy population ($\sim$10\%) is found to be extremely compact ($<$0.1 kpc), barely resolved in NIRCam images. These results indicate that large star formation rate surface densities ($\Sigma_{\rm{SFR}}$)  become increasingly common at high redshifts \citep{Ono2023,Morishita2024,Calabro2024}, potentially resulting in somewhat different galaxy formation conditions at very high redshifts. This may enhance the fraction of star formation in dense bound clusters, as discussed below, and it may also contribute to formation of the high N/O abundance patterns (\S\ref{sec:abundances}) and a population of very massive or super-massive stars (\S\ref{sec:highionization}). 

Galaxies exhibit relationships between both stellar mass and size and UV luminosity and size (Fig.~ 7a). At $z\simeq 4.5-5.5$, \citet{Ono2024} find median half-light radii increase from 0.36 kpc at M$_{\rm{UV}}=-19.89$ to 0.77 kpc at M$_{\rm{UV}}=-20.95$. The (rest-UV) size mass relationship  is described by a power law (R$_e \propto \rm{M}_\star^{0.19}$), with a similar slope as that seen at $z\simeq 3$ \citep{Morishita2024}. 
However when viewed at fixed luminosity or mass, NIRCam-measured galaxy sizes evolve significantly over $0<z<10$ \citep{Ono2023,Morishita2024, Ormerod2024}.  Constraints on the power law redshift evolution of the characteristic sizes  (1+z$^{\alpha}$) have so far shown similar power law indices as were found previously with {\it HST} ($\alpha \sim -1)$.  When the comparison is limited to high redshift ($5<z<14$), the evolution is slower, with a power law index of $\alpha \sim -0.4$ \citep{Morishita2024,Sun2024}, consistent with findings over similar redshift ranges with {\it HST} \citep{CurtisLake2016}. Comparison of {\it JWST} size measurements to simulations is ongoing, with several simulations finding median apparent sizes that are significantly larger than what is observed  in M$_\star=10^6$ to 10$^9$ M$_\odot$ galaxies at $z\gtrsim 6$ \citep{Shen2024}, and other simulations finding compact ($<$100 pc) galaxies are common at $z\gtrsim 5$ \citep{Pallottini2022}. These differences may reflect variations in feedback prescriptions \citep{Shen2024}.

{\it JWST} also has provided the first spatially-resolved view of the stellar population ages within $z\gtrsim 5$ galaxies. At lower redshifts ($z\simeq 1.5-2.5$), galaxies look very different in the rest-UV and rest-optical. Blue clumpy star forming structures are often found surrounding  a centrally-located red stellar component 
\citep[e.g.,][]{Wuyts2012, Lee2013}. The UV light is clumpy, but the longer-wavelength light reveals a smoother distribution centered on the red nuclear component. This picture reflects the inside-out growth of galaxies, with the outer blue regions actively building up at $z\simeq 1.5-2.5$.  Prior to {\it JWST}, it was not known whether the UV-emitting clumps (seen with {\it HST}) were co-spatial with rest-optical light (seen with {\it Spitzer}), or whether they are embedded or at the outskirts of a larger distribution of older stars. NIRCam imaging delivered the first high resolution view of the rest-frame optical light at $z\gtrsim 5$. Many brighter $z\gtrsim 5$ galaxies were found to be clumpy (as was seen with {\it HST}), with more than half of the light coming from  10$^7$ M$_\odot$ and 10$^9$ M$_\odot$ star forming complexes that are 100-500 pc in size (Fig~ 7b). The stellar population properties vary from clump to clump within individual galaxies, but the majority of clumps are blue and dominated by young stars. Unlike galaxies at lower redshifts, the rest-optical continuum has been found to be just as clumpy as the rest-UV, and in many galaxies no evidence for a centrally-located older stellar component has been identified. The rest-UV sizes are very similar to those in the rest-optical  \citep{Yang2022, Chen2023, Ono2024, Sun2024}.  This suggests that the dominant mass component in $z\gtrsim 5$ galaxies is plausibly contained in the clumpy star forming complexes, a picture that would be expected if many $z\gtrsim 5$ galaxies are still experiencing the first phase of inside-out growth at these redshifts. A similar picture is suggested by spatially-resolved measurements of H$\alpha$, with galaxies appearing to undergo rapid build-up of compact bulges at $z\gtrsim 5$, transitioning to inside-out growth of the disk at lower redshifts \citep{Matharu2024}. 
There is a subset of galaxies at $z\gtrsim 5$ that appear to show more evolved structures.
\citet{Baker2024} present a $z=7.4$ galaxy with a core with stellar mass densities approaching those seen in local massive elliptical galaxies, surrounded by more extended star forming complexes. \citet{Nelson2024} find an extended rotating galaxy with centrally depressed star formation at $z=5.3$, consistent with a central bulge surrounded by a young disk as is often seen at $z\simeq 2$. As more examples of galaxies with mature structures are discovered, it will become possible to better understand the conditions under which they form at such early epochs.

Strong gravitational lensing has extended our view of galaxy sizes and structures to extremely small scales. When viewed with magnification factors of 10-100$\times$, galaxies still appear very clumpy, with the majority of the luminosity emerging from dense star cluster complexes. In several recent lensed examples, $z\gtrsim 6$ galaxies have been resolved into more than ten high surface brightness $\sim 1$-50 pc star cluster complexes \citep{Vanzella2023, Fujimoto2024, Mowla2024, Adamo2024, Vanzella2024}. 
These compact and dense structures are found with a range of stellar masses (10$^4$-10$^7$ M$_\odot$) and often extremely large stellar surface densities ($\gtrsim 10^5$ M$_\odot$ pc$^{-2}$), three orders of magnitude larger than what is seen in young star clusters at the present day. The inferred dynamical ages suggest that the star clusters are likely to be gravitationally-bound systems. Those that survive until the present-day may be progenitors of the massive metal poor globular clusters seen in nearby galaxies. When dense clusters are observed at extremely young ages ($\lesssim$1-3 Myr), they appear to show an N/O abundance pattern that is similar to that seen in the second generation stellar populations of globular clusters (\S\ref{sec:abundances}, \citealt{Topping2024b}). Overall these results suggest that much of the star formation in early galaxies may be occurring in bound star clusters, possibly reflecting the very dense gas-rich conditions that appear common at $z\gtrsim 5$ \citep[e.g.,][]{Ma2019}. It has been suggested that such dense star clusters may be ideal environments for rapid formation and growth of supermassive black holes \citep[e.g.,][]{Shi2024}, perhaps contributing to the detections of (potentially) massive black holes at $z\gtrsim 5$ with {\it JWST} (\S\ref{sec:AGN}).

Attention has also focused on the morphological and structural parameters. The results have demonstrated that galaxies with disk-like light profiles are surprisingly common at very high redshift, comprising upwards of 30-50\% of the $z>4$ population \citep[e.g.,][]{Ferreira2023, Robertson2023, Ormerod2024, Nelson2023, Kartaltepe2023, Sun2024, Conselice2024, HuertasCompany2024, Lee2024, Tohill2024}. Integral field spectroscopy of gas kinematics will be required to confirm that the disk-like profiles at very high redshifts  correspond to rotation-supported structures \citep[e.g.,][]{Wang2024}. Observations with ALMA have demonstrated that velocity gradients are present in some early galaxies \citep[e.g.,][]{Smit2018}, and {\it JWST} measurements of ionized gas kinematics are starting to build on these detections \citep[e.g.,][]{Nelson2023, Xu2024, deGraaff2024}. Statistical samples will be required to map the fraction of $z\gtrsim 5$ galaxies showing  hints of rotation.

\section{Gas, Dust, Metals}\label{sec:ism}

The ionized gas properties are most effectively probed in the rest-frame optical. In the last two years, {\it JWST} has provided our first spectroscopic measures of the rest-optical emission line properties of $z\gtrsim 6$ galaxies. Individual galaxy spectra will often show a subset of strong emission lines ([OII], [Ne III], H$\beta$, [OIII]), with fainter lines visible in deeper exposures (and H$\alpha$ visible up to $z\simeq 6$). Analyses of emission line properties currently focus on samples of up to 2400(700)  and $z\gtrsim 4$(6), allowing statistical investigations of ionized gas and star formation.
These numbers will grow considerably in the months to come. In this section, we provide a brief review of what these data are telling us about the ionization conditions, gas-phase metallicity, electron density, and abundance patterns  in early galaxies.

\subsection{Ionization Conditions}\label{sec:ionization}

The gas ionization state is often parameterized empirically by flux ratios of emission lines probing gas in two adjacent ionization states. Most commonly used are the flux ratios of [OIII] and [OII] (O32) and [Ne III] and [OII] (Ne3O2).  These ratios offer empirical proxies for the ionization parameter (U), the ratio of the number density of ionizing photons and number density of hydrogen atoms. For ionization-bounded nebulae, the ionization parameter depends on the ionizing photon rate (Q), the electron density (n$_{\rm{e}}$), and filling factor of ionized gas ($\epsilon$). For spherical nebulae, it is straightforward to show that the ionization parameter goes as $U\propto [Q~$n$_{\rm{e}}$ $\epsilon^2$]$^{1/3}$. Hence variations in the ionization-sensitive proxies (O32 and Ne3O2) may reflect changes in both ionization and gas conditions. 

It has been shown in early {\it JWST} spectroscopy that O32 and Ne3O2 tend to be large in many reionization-era galaxies. As was previously found at $z\simeq 2$ \citep{Tang2019, Sanders2020}, the  ionization-sensitive ratios  correlate with rest-optical EWs, varying from O32=1 (Ne3O2=0.1) at [OIII]5007 EW = 200~\AA\ to O32=20 (Ne3O2=1.5) at [OIII]5007 EW = 2000~\AA\  \citep{Tang2023,Boyett2024}. Given the connection between [OIII] EW and sSFR (\S\ref{sec:galaxyprops}), this result  suggests a physical link between the sSFR and the ionization parameter of the gas. This may reflect enhancements in gas density, gas filling factor, and ionizing photon rate  during  strong bursts that power large sSFR galaxies. \citet{Reddy2023} demonstrate that the ionization parameter is also correlated with the star formation rate surface density in high redshift galaxies.  They show that the O32 trends appear to be primarily driven by variations in the electron density (and not ionizing photon rate), with larger n$_{\rm{e}}$ typical in galaxies with larger sSFR and $\Sigma_{\rm{SFR}}$.  

The $z\gtrsim 6$  population includes galaxies with extremely large rest-optical EWs that have rarely been seen at lower redshifts. In these galaxies, the O32 and Ne3O2 ratios also approach values that have rarely been observed. For example,  the galaxy RXCJ2248-ID (a compact burst with [OIII] EW = 2800~\AA) has  O32 = 184 and Ne3O2 = 16 \citep{Topping2024b}. Galaxies with such large [OIII]+H$\beta$ EWs and ionization-sensitive ratios tend to have extremely large electron densities (10$^5$ cm$^{-3}$) where [OII] is significantly collisionally de-excited. In these cases, the large O32 values are driven  by large ionization parameter and collisional de-excitation of [OII], albeit with both  related to the large densities associated with strong bursts.

\subsection{Metallicity}\label{sec:metallicity}

The gas-phase metallicity of a galaxy is regulated by the interplay of star formation, outflows, and accretion of gas. The standard approach to  measurement of metallicity  is through observations of rest-frame optical emission lines in galaxy spectra, either via  the direct method when faint auroral lines are detected or via strong-line calibrations. Over a decade ago, large near-infrared spectroscopic surveys led to the first statistical samples of gas-phase metallicities at $z\simeq 2$ \citep[e.g.,][]{Kashino2013, Steidel2014, Kriek2015}, providing our first window on the relationship between stellar mass and metallicity (MZR) in $z\sim 2$ galaxies. 
At $z\gtrsim 4$, the strong rest-optical lines required for metallicity inferences are redshifted out of the wavelength window that ground-based facilities are able to probe. As a result, little was known about the typical metallicities of galaxies in the reionization era. 
This changed immediately with the launch of {\it JWST}. The sensitivity of NIRSpec at observed wavelengths of 2-5$\mu$m is optimized for detecting  rest-frame optical emission lines at $z\gtrsim 4$. In particular, the [OIII]$\lambda$4363 auroral line can now be recovered in large samples out to $z\simeq 10$, enabling direct method metallicities across a much wider redshift range than was possible prior to {\it JWST} and providing improved calibrations for strong line metallicities \citep{Sanders2024Te}.  At $z\simeq 2-4$, the sensitivity of {\it JWST} allows multiple auroral lines to be detected, enabling the abundances of multiple atomic species in various ionization states to be constrained \citep[e.g.,][]{Rogers2024}. The relative abundance ratios will enable more detailed constraints on chemical evolution, tracking elements that build up in the ISM on different timescales (see \S\ref{sec:abundances}). 

In the last two years, {\it JWST} has provided the first constraints on the MZR at $z\gtrsim 4$ \citep[e.g.,][]{Nakajima2023, Curti2024, Morishita2024mzr, Sarkar2024, Chemerynska2024, RobertsBorsani2024}. Initial work has used both auroral line samples and  strong line calibrations, with samples sizes of $80-150$ galaxies at redshifts $z>3$. At $3<z<6$, these studies generally find metallicities increase from  $12+\log(\rm O/H)=7.55$ (Z=$0.07~Z_{\odot}$) at $10^7~\rm M_{\odot}$ to $\log(\rm O/H)=7.87$ (Z=$0.15~Z_{\odot}$) at $10^9~\rm M_{\odot}$ (Fig.~8a). 
Many of these  studies show only mild evolution in the MZR at higher redshift.  For example, \citet{Curti2024} derive metallicities at $z=6-10$ that are only $0.15$ dex lower than at $z=3-6$ in the high-mass ($>10^8~M_\odot$) end of their sample, while no evolution is seen at lower masses. Similarly, \citet{Sarkar2024} and \citet{Nakajima2023} also report very little change in the MZR at $z\gtrsim 3$.  
The slow evolution in the MZR at $z\gtrsim 3$ stands in contrast from the more pronounced evolution in normalization ($\sim 0.5$ dex) seen between $z\simeq 0$ and $z\simeq 3$.  Early estimates of the MZR scatter have found values ($0.1-0.15$ dex) that are in broad agreement with the scatter seen at $z=0-3$ \citep{Curti2024, Sarkar2024}. There are some indications from gravitationally lensed galaxies that the scatter may be somewhat larger in low mass galaxies \citep{Chemerynska2024}. 
The MZR observations are now being directly compared to predictions from simulations.  While some  do predict that there should be modest (0.2-0.3 dex) evolution in the MZR at $z\simeq 4-8$ \citep{Torrey2019,Wilkins2022}, 
others find the evolution to be negligible \citep{Langan2020,Ucci2023,Marszewski2024}, consistent with the observational trends. Some simulations predict a large scatter in the MZR at low masses \citep{Marszewski2024}, possibly reflecting the bursty nature of star formation. 
While it is not clear that this is supported by current observations \citep{Pallottini2024}, the scatter is  challenging to robustly measure. As we noted in \S\ref{sec:galaxyprops}, 
continuum-selected surveys often do not produce representative samples at the lowest masses  (see also \citealt{Sun2023completeness}), while additionally  the lowest metallicity galaxies may be missed in flux-limited surveys. Deeper spectroscopy  should better clarify the scatter in the low mass MZR, providing another probe of the variability of SFR in high redshift galaxies.

\begin{figure}[t]
\centering
\includegraphics[width=.9\textwidth]{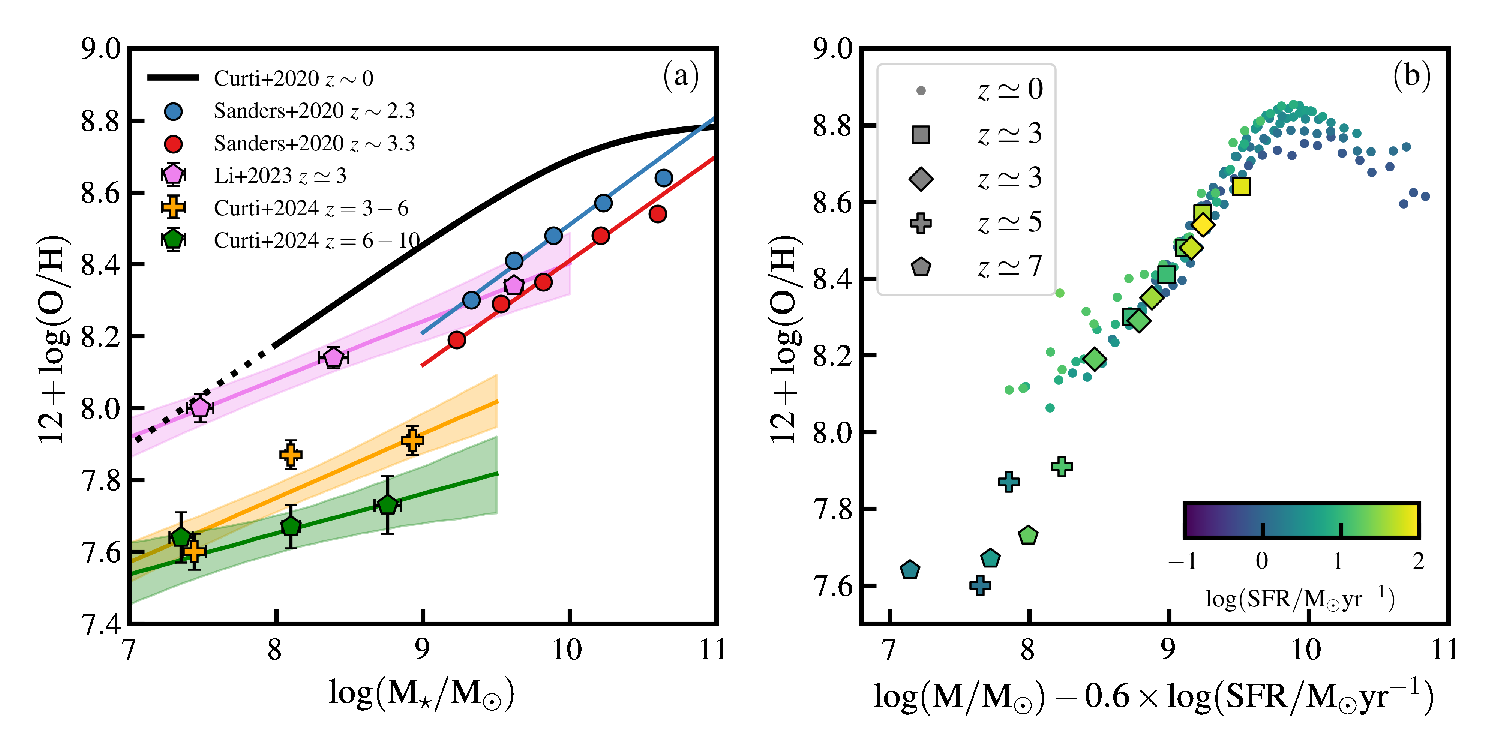}
\caption{(a) Mass metallicity relationship at $z\gtrsim 3$ from {\it JWST}, revealing typical gas-phase metallicities of $\sim 0.1$ Z$_\odot$ at $z\gtrsim 5$. Current results suggest that the mass metallicity relationship does not evolve strongly at $3<z<10$, in contrast to the significant shift toward lower metallicities that appears between $z\simeq 0$ and $z\simeq 3$.  (b) Fundamental metallicity relation (FMR) at $z\simeq 0-7$, combining metallicity, stellar mass, and star formation rate. {\it JWST}  indicates that there may be a deviation from the FMR among low metallicity $z\gtrsim 5$ galaxies. }
\label{fig:MZR}
\end{figure}

Galaxies at $z\simeq 0-3$  show a secondary dependence of the metallicity on the SFR \citep[e.g.,][]{Ellison2008, Mannucci2010, Cresci2019, Curti2020, Sanders2021}. The mass-Z-SFR relation (i.e., Fundamental Metallicity Relation; FMR) is thought to arise  as pristine gas inflows dilute the ISM metallicity while simultaneously boosting the SFR \citep[e.g.,][]{Peeples2011}. In this framework, higher SFR at fixed mass implies lower metallicity.
 Characterization of the FMR with {\it JWST} at $z\gtrsim4$ shows a potential deviation from the low-redshift FMR toward lower metallicity (Fig.~\ref{fig:MZR}b). If confirmed, this may indicate that physical processes that dilute the ISM (e.g., pristine gas accretion or feedback) may be enhanced at early times \citep{Curti2024}. Some simulations similarly predict that galaxies become increasingly offset toward lower metallicities as redshift increases, in agreement with current observations \citep{Sanders2023b, Garcia2024}.

In addition to the statistical investigations described above, there is great interest in identifying high redshift galaxies with near-pristine metallicities. 
In nearby star forming galaxies, there is an apparent gas-phase metallicity floor such that oxygen abundances approaching $12+\log(\rm O/H)=6.69$ ($0.01Z_{\odot}$) become exceedingly rare \citep{Sanchez-Almeida2017,Senchyna2019}.  
{\it JWST} has extended the search for metal-free galaxies to very high redshift, where young and low mass galaxies are common. It has long been predicted that metal-free pockets may persist until lower redshifts in voids \citep[e.g.,][]{Scannapieco2003,Trenti2009, Pallottini2014}.
The key signatures are very strong hydrogen recombination lines coupled with extremely weak metal emission lines. 
If Pop III stars with a top-heavy IMF are present, we may additionally expect strong He II emission. 
Existing calibrations between strong emission line ratios and metallicity predict that [OIII]$\lambda5007$ becomes weaker than H$\beta$ below $12+\log(\rm O/H)\simeq6.69$ ($\simeq0.01~Z_{\odot}$).
In spite of the large spectroscopic samples that have been obtained in the last two years, we have yet to find many spectra that push the metallicity floor below what has been seen in nearby galaxies, with few systems showing [OIII]/H$\beta<1$ \citep{Morishita2024mzr}. This may simply reflect the rapid enrichment in oxygen from core-collapse supernovae, or the larger stellar masses we have mostly probed thus far.
One of the only known near-pristine galaxies identified to date was found via gravitational lensing at $z=6.639$
\citep{Vanzella2023}. Previously detected only by its Ly$\alpha$ emission from MUSE, observations with {\it JWST} demonstrate a host of strong hydrogen Balmer lines in addition to much weaker emission from oxygen lines. The observed $\rm[OIII]/H\beta$ ratio of 0.55 indicates a very low metallicity of just $12+\log(\rm O/H)\lesssim6.3$ ($\lesssim 0.004 ~ Z_{\odot}$), making it the most metal-poor galaxy known at high redshift. As spectroscopy extends to ultra low mass galaxies in lensing fields ($\lesssim 10^5$ M$_\odot$), we may find more examples that push the low metallicity floor further toward the metal free regime.

\subsection{Ionized Gas Densities}\label{sec:density}

The ionized gas density can be measured using  density-sensitive doublet ratios. The low ionization gas in galaxies can be constrained in the rest-optical by [OII] and [SII], while the rest-UV doublet ratios of Si III], C III], and N IV]  (ionization energies of 16.3 eV, 24.4 eV, and 47.4 eV, respectively) probe gas at moderate and high ionization conditions. HII regions  and galaxies will often have density gradients \citep[e.g.,][]{Kewley2019}, with the highest ionization gas tracing the highest densities. Additionally there may simply be multiple density components within individual galaxies. Observationally higher ionization transitions at lower redshifts do tend to trace higher density gas \citep{James2014, Mainali2023}. This trend is also seen in some simulations \citep{Katz2023}. Knowledge of the range of ionized gas densities is critical for robust interpretation of emission line spectra, while additionally offering unique insight into the gas conditions.

The sensitivity of NIRSpec enables the first detailed exploration of ionized gas densities in high redshift galaxies. One of the first surprises  was the observation that a subset of $z\gtrsim 6$ galaxies host ionized gas with much larger electron densities  than are seen (in lower ionization gas) at $z\simeq 2-3$. This was first noticed in the $z=10.6$ galaxy GNz11 \citep{Bunker2023}, where the limit on the flux ratio of the NIV] doublet pointed to gas with $\gtrsim 10^5$ cm$^{-3}$ \citep{Senchyna2024,Maiolino2024}. With GNz11, it has been argued that the high electron densities may reflect the broad line region of an AGN \citep{Maiolino2024}. Since then, several additional examples of high density gas ($\simeq 10^5$ cm$^{-3}$) have been uncovered with more robustly-measured density doublet ratios  and no clear evidence for AGN activity.  Physically the elevated gas densities  appear to arise primarily in the  subset of  galaxies experiencing the strongest bursts of star formation, where [OIII]+H$\beta$ EWs are in excess of 2500~\AA. It is plausible these measurements reflect the high gas densities associated with the large sSFR phases. 
At very high density, the spectra are significantly altered in several ways. Many lower ionization lines (i.e., [OII]) will be collisionally de-excited (see discussion in \S\ref{sec:ionization}). The He I$\lambda$5877 emission line will appear atypically strong (up to 25\% of the H$\beta$ flux), reflecting the enhanced collisional excitation rates. In cases where only low resolution spectra are available, density-sensitive doublets are not resolved. If extremely large  O32 or large He I$\lambda$5877/H$\beta$ ratios are present, it may provide a signpost of very high density conditions.

There have also been  statistical explorations of electron densities in high redshift galaxies. The most commonly-used indicators  have typically been the rest-optical [OII] and [SII] doublets. Prior to {\it JWST}, it was shown 
that the average electron density increases from 
$\sim 30/\rm cm^{3}$ at $z\sim0$ \citep[e.g.,][]{Davies2021} to $\sim250/\rm cm^{3}$ at $z\simeq2-3$ \citep[e.g.,][]{Sanders2016}.  {\it JWST} has made it possible to expand these constraints to $z\gtrsim4$ with the wavelength coverage of NIRSpec at $2.5-5.0~\rm \mu m$.
Existing samples of [OII] and [SII] detections at $z\simeq 4-9$ reveal densities of $\gtrsim300/\rm cm^{3}$, representing an increase over values at $z\simeq 1-3$ for a fixed stellar mass and SFR \citep[e.g.,][]{Isobe2023}.
At yet higher redshift, densities for the brightest galaxies are being constrained, with one of the best-studied systems being MACS0647-JD at $z\sim10.167$. High-resolution {\it JWST}/NIRSpec spectroscopy of the [OII] doublet reveals a doublet ratio that implies an electron density of $\sim800/\rm cm^{3}$ \citep{Abduroff2024}.
The combination of  measurements over $0<z<10$ show an average evolution at a rate of $\propto(1+z)^{1-2}$ \citep{Isobe2023,Abduroff2024}. The precise physics driving the evolution in the ionized gas densities is not clear, but the observed trend may reflect lower metallicities and smaller galaxy sizes at earlier times.  

\subsection{Dust: Far-IR Continuum, Attenuation Curves, 2175 Angstrom Bump}\label{sec:dustprops}

Over the last 5-10 years, our understanding of the ISM in early galaxies has developed rapidly, thanks in large part to observations of the rest-frame FIR with ALMA. Whereas rest-UV continuum slopes (see \S\ref{sec:uvslopes}) are impacted by the reddening effects of dust, the rest-FIR probes dust directly through its continuum emission. We will briefly describe recent results, focusing on UV-detected galaxies (and not dusty sub-millimeter systems). More detailed discussion  can be found in recent reviews \citep[e.g.,][]{Hodge2020}. 

Two large ALMA programs have greatly increased the number of dust continuum detections at high redshift.
The ALMA Large Program to INvestigate [CII] at Early times (ALPINE) focused on dust in galaxies at $z\simeq 5-6$ \citep{Khusanova2021}, and the Reionization Era Bright Emission LIne Survey (REBELS, \citealt{Bouwens2022}) extended this to $z>6.5$ galaxies. Both surveys observed UV-luminous continuum-selected galaxies identified over wide areas.  In particular, REBELS observed 40 UV-bright galaxies selected in 7 deg$^2$ of imaging, using spectral scans to confirm redshifts with [CII]$\lambda$158$\mu$m, [OIII]$\lambda$88$\mu$m, while simultaneously observing the dust continuum emission, building on strategies developed in earlier pilot programs \citep{Smit2018, Schouws2022}. Combined, REBELS provides observations of 49 UV-selected galaxies at $z>6.5$. \citet{Inami2022} present dust continuum measurements for 18 of the 49 galaxies, tripling the number of known detections at $z>6.5$. The inferred infrared (IR) luminosities (L$_{\rm{IR}}$) range from 3 to 8$\times$10$^{11}$ L$_\odot$. Only one source (REBELS-25, discussed in \citealt{Hygate2023}) has an IR luminosity (1.5$\times$10$^{12}$ L$_\odot$)  that classifies it as an ultra-luminous infrared galaxy (ULIRG, 10$^{12}$ $<$ L$_{\rm{IR}}$/L$_\odot$ $<$  10$^{13}$). The fraction of obscured star formation in the REBELS galaxies typically range from 50 to 90\% \citep{Inami2022, Ferrara2022, Bowler2022, Schouws2022}, similar to the range found in $z\simeq 5$ galaxies in ALPINE \citep{Fudamoto2020}. Dust masses have been estimated for REBELS galaxies using methods described in \citet{Sommovigo2022}, with typical inferred values of 10$^7$ M$_\odot$.  The  dust-to-stellar-mass ratio in REBELS galaxies faces uncertainty owing to the outshining problem (\S\ref{sec:galaxyprops}), but most values appear broadly consistent with predicted ratios ($\sim0.1$~\%) in simulations \citep{Dayal2022, Ferrara2022, Narayanan2024}. 
Spatial offsets between the UV and dust continuum ($\sim 0.5$-1.5 arcsec) are found in some of the REBELS galaxies. This suggests a decoupling between the components dominating the UV light and the dust continuum, as is often seen in sub-millimeter galaxies. This is broadly consistent with the {\it JWST} finding that the most UV luminous $z\gtrsim 6$ galaxies tend to be comprised of multiple clumps, each with a range of stellar population ages and reddening (see \S\ref{sec:sizes}, Fig~\ref{fig:sizes}b). The REBELS galaxies may represent a higher mass extension of the luminous $z\simeq 6-7$ galaxies identified in the existing {\it JWST} deep fields. 

The ALMA dust continuum measurements have allowed for the first statistical investigations of the  relationship between UV slope and infrared excess (IRX=log$_{10}$ (L$_{\rm{FIR}}$/L$_{\rm{UV}}$)) in  $z\gtrsim 5$ galaxies. 
The relationship between IRX and $\beta$  depends on the attenuation law, which in turn is sensitive to the dust grain properties  and the geometric distribution of dust and stars (see \citealt{Salim2020}). For local starburst galaxies, the "IRX-$\beta$ relation was presented in \citet{Meurer1999} and \citet{Calzetti2000}. Whether this relation holds at higher redshift has been the subject of considerable work in recent years. \citet{Bowler2024} investigated the IRX-$\beta$ in 49 UV-bright galaxies identified in the REBELS survey, finding results that are consistent with the Calzetti local starburst relation. When folding in the ALPINE sample, they demonstrate little evolution in the IRX-$\beta$ relationship for UV-bright galaxies over $4<z<8$. This is consistent with most predictions from simulations \citep[e.g.,][]{Narayanan2018, Liang2019, Pallottini2022, Vijayan2024} and semi-analytic models \citep{Mauerhofer2023}.
\citet{Bowler2024} compare the redshift evolution of IRX at fixed stellar mass in the combined ALPINE and REBELS samples, finding evidence for a significant ($>$3$\times$) decrease in obscured star formation at $z>4$ compared to those derived at $z\simeq 3$. 
 
Prior to JWST, efforts to directly constrain the nebular attenuation curve beyond the local Universe have been limited. Even with state-of-the-art facilities, the combined sensitivity and wavelength range made it possible to only derive an average curve at $z\simeq2$ from just the few brightest Balmer lines \citep{Reddy2020}. Nevertheless, this campaign demonstrated that the average nebular attenuation curve closely matched the Milky Way extinction curve \citep{Cardelli1989} at optical wavelengths. In contrast, the stellar attenuation curve received extensive characterization based on the shape of galaxy SEDs or UV continuum slopes combined with FIR emission, in which considerable variation was seen among individual $z\simeq2$ galaxies \citep{Salmon2016, McLure2018, Battisti2022}.
\citet{Shivaei2020} demonstrated that the average stellar attenuation curve of metal-rich ($12+\log(\rm{O/H})>8.5$) $z\simeq2$ galaxies closely matched that that of local starbursts \citep{Calzetti2000}, while the attenuation curve of their metal-poor counterparts was better matched by the SMC extinction curve \citep{Gordon2003}.
The landscape of attenuation at high redshift has widened substantially with launch of JWST. The improved sensitivity provides the necessary constraints to derive nebular attenuation curves for individual galaxies, while coverage at $>3\mu m$ enables the detection of the brightest Balmer lines at $z>4$ for the first time. Theses capabilities have been realized with the first nebular attenuation curve at $z>4$ that is markedly different than either $z\simeq2$ or local galaxies \citep{Sanders2024}. While bespoke attenuation curves will ultimately lead to more precise galaxy properties, such methodology is still in its infancy as it has not yet been applied to typical high-redshift galaxies. However, it will soon become routine for analyzing the emission of early galaxies.

{\it JWST}  is beginning to offer direct probes of dust in $z>5$ galaxies. \citet{Witstok2023}
presented spectroscopic detection of the 2175~\AA\ dust attenuation bump in a $z=6.71$ galaxy. This feature is thought to be due to excess absorption from small carbonaceous dust grains. The UV bump is seen in both the extinction curves of the Milky Way and LMC, but it is weak in that of the SMC. The presence or absence of the feature depends on the grain size distribution (specifically the presence of very small grains), which in turn is sensitive to the star formation history. Previous work has identified the 2175~\AA\ bump in high mass dusty galaxies at $z\simeq 3$ \citep[e.g.,][]{Shivaei2022}. The detection of a UV bump in a lower mass galaxy at $z\gtrsim 6$ is surprising in this context, potentially suggesting differences in the grains driving the absorption feature. The detection of the UV bump requires a rapid production of carbon and subsequent depletion onto the grains responsible for the absorption features. Most existing measurements suggest early galaxies have sub-solar C/O ratios (see \S\ref{sec:abundances}). Those galaxies with UV bumps may correspond to outliers with larger C/O ratios, or systems with SNe or WR stars that are efficient at producing carbonaceous grains as discussed in \citet{Witstok2023}. Larger spectroscopic samples of UV bumps at high redshift will be required to confirm how frequently these conditions exist in early galaxies.  If UV bumps are found to be common in lower mass galaxies at $z\gtrsim 7$, it would place strong constraints on dust formation mechanisms at very high redshift.

\subsection{Abundance Patterns: N/O and C/O Ratios}\label{sec:abundances}

In addition to constraining the metallicity, deep {\it JWST} spectra provide insight into the chemical abundance pattern in early galaxies. Different elements will pollute the ISM on different timescales, so measurement of their relative abundance offers insight into the past star formation history.  The alpha-elements (i.e. oxygen) will appear shortly after a burst of star formation from core-collapse supernovae. Other elements will increase in abundance on longer timescales, once the metal-rich winds of evolved intermediate-mass stars begin to pollute the ISM. 
In particular, the winds of asymptotic giant branch (AGB) stars are expected to increase the abundance of carbon and nitrogen on timescales of $\simeq 100$ Myr after a burst of star formation. As a result, both carbon-to-oxygen ratios  (C/O) and nitrogen-to-oxygen ratios (N/O) are  sensitive to the recent star formation history, with potential to reveal whether or not the stellar population is dominated by stars younger or older than $\simeq 100$ Myr. 
In metal poor galaxies (i.e., 12+log O/H $<$ 8.0) locally and at lower redshifts,  C/O and N/O ratios tend to be considerably lower than  the solar values (log (C/O)$_\odot$=-0.23 and $\rm{log (N/O)_\odot}=-0.86$). The relative carbon abundance plateaus to log (C/O) $\simeq -0.7$ at these low metallicities, while the nitrogen abundance is generally found to be  log (N/O) $\lesssim -0.5$ in similarly metal poor systems. At extremely low metallicities ($<$1\% Z$_\odot$), the C/O ratio may rise to higher values, reaching close to the solar C/O ratio at [O/H] = -2.5 to -3 \citep[e.g.,][]{Cooke2011}. This may reflect 
carbon-enhanced yields of the first generation of stars.

Measurements of $z\gtrsim 6$ abundance patterns with {\it JWST} have mostly focused on C/O and N/O ratios thus far (Fig.~9). Determination of robust C/O ratios typically requires detection of the [CIII], CIII] and OIII] UV doublets \citep[e.g.,][]{Berg2019b}. In cases where both UV lines have been detected with {\it JWST} in metal poor galaxies at $6<z<13$, the derived C/O ratios have all been found to be sub-solar, consistent with expectations for delayed carbon production in young, metal poor galaxies. There has been one galaxy at $z\simeq 12$ (GS-z12) for which a near-solar value of C/O has been claimed \citep{DEugenio2024}. However this measurement is made combining CIII] and [OII] emission lines. As the authors note, if the C/O ratio is instead calculated in the standard manner (i.e. with CIII] and OIII]), the C/O value in this galaxy is found to be sub-solar. 

The N/O ratios have also generated great interest.
Historically N/O has been mostly derived using rest-optical tracers (i.e., [NII] and [OII]). At $z\gtrsim 6$, the first measurements have focused on the higher ionization lines in the rest-frame UV (NIV], [NIII]).  In one of the more surprising results,  N/O  values have been shown to take on near-solar ratios in a subset of metal poor galaxies at $z\gtrsim 6$, an abundance pattern that is very distinct from what is common in galaxies at later cosmic epochs.
This was first seen in the spectrum of the $z=10.6$ galaxy GNz11 \citep{Bunker2023}, based on detections of the NIV]$\lambda$1483,1486 doublet and NIII] quintuplet centered at 1750~\AA.  Similar nitrogen lines and N/O abundance patterns have now been reported robustly in 5 other metal-poor galaxies at $z\gtrsim 6$ \citep[e.g.,][]{Topping2024b, Topping2024c, Castellano2024, Schaerer2024}. The demographics of this population are becoming more clear, with current data suggesting that elevated N/O ratios are primarily limited to galaxies undergoing compact bursts of star formation with [OIII]+H$\beta$ EWs in excess of 2500~\AA. This is the same subset of the galaxy that appears to produce the hard ionizing spectra described in \S9 and high electron densities discussed in \S\ref{sec:density}. Some nitrogen emitters do show evidence of AGN \citep[e.g.,][]{Ji2024}, but existing data does not indicate that this is a necessary criteria for nitrogen enhancement.

This nitrogen-enhanced abundance pattern suggests that a completely distinct chemical evolution pathway is present in a subset of $z\gtrsim 6$ galaxies. While almost never seen in lower redshift galaxies (though see \citealt{Fosbury2003, McGreer2018,Pascale2023}), this abundance pattern is present in nearly all massive globular clusters \citep[e.g.,][]{Gratton2004,Bastian2018}. The abundances are consistent with the yields of high temperature nuclear burning. How this enriched material is ejected into the ISM and locked into stars without being dominated by the more common yields of supernovae is not well-established. One of the more successful scenarios involves the winds of very massive stars ($>$100 M$_\odot$) or super-massive stars ($>$1000 M$_\odot$) \citep[e.g.,][]{Nagele2023, Cameron2023, Charbonnel2023, Senchyna2024}. How such stars are formed is not clear, but it has been proposed that they could be assembled up from stellar collisions or runaway accretion, 
both of which may be expected in the dense environments where globular clusters form.  The {\it JWST} discovery of anomalous N/O ratio at very high redshift allows direct searches for these proposed extreme stellar populations. The stellar densities indeed appear extremely high in the most magnified examples \citep{Topping2024a}, but current information on the nature of the massive stars in these galaxies remains limited. Deep continuum spectroscopy should plausibly reveal strong P-Cygni wind signatures (i.e., NV, CIV) from very massive stars if they are the primary polluters responsible for the anomalous abundance pattern. 

\begin{figure}[t]
\centering
\includegraphics[width=.9\textwidth]{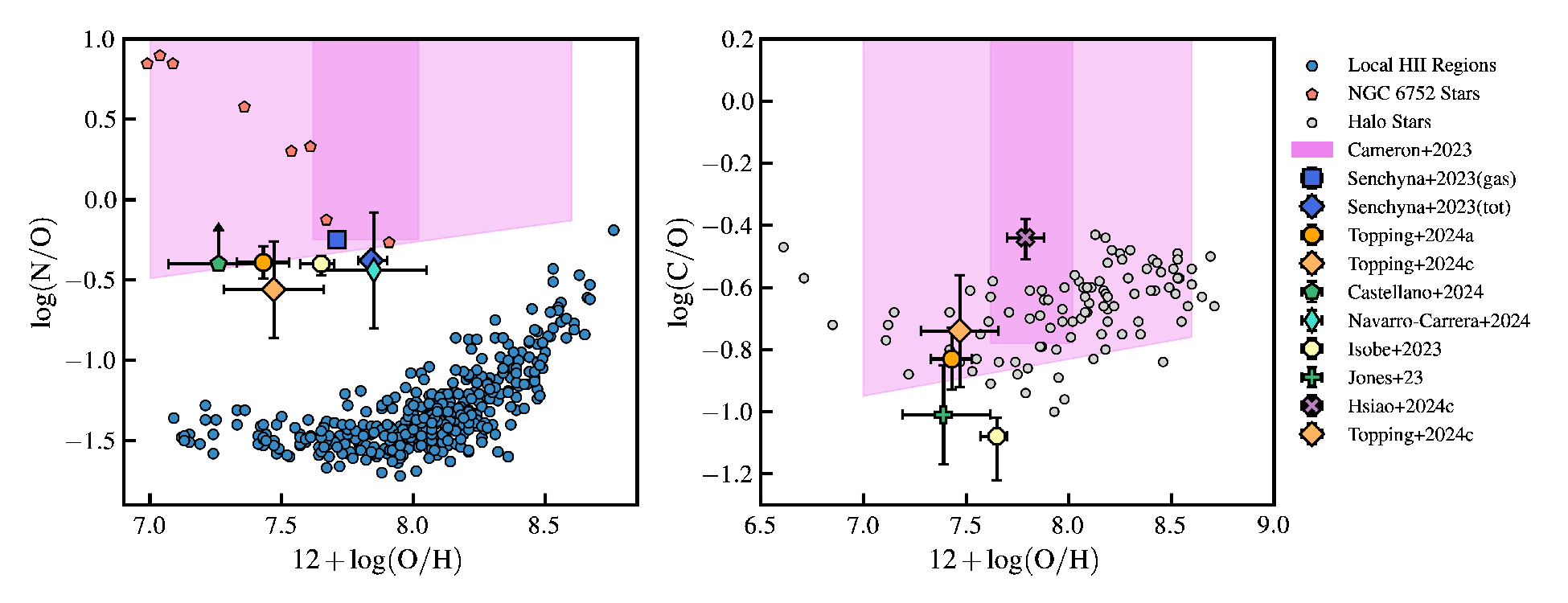}
\caption{Chemical abundance pattern in $z\gtrsim 6$ galaxies. (a) N/O vs O/H for $z\gtrsim 6$ galaxies with rest-UV nitrogen line detections. The N/O values are well above what is seen in nearby metal poor galaxies, but are similar to what is found in many globular cluster stars. The stellar origin of the elevated N/O is not known. (b) C/O vs O/H for galaxies with rest-UV OIII] and CIII] detections. Existing measurements indicate sub-solar C/O ratios, consistent with what is seen in nearby metal poor galaxies.  }
\label{fig:CNO}
\end{figure}

\section{Nature of Early Ionizing Sources: Implications of High Ionization Lines in Rest-UV Spectra}\label{sec:highionization}

The rest-frame UV is host to a suite of emission lines from highly-ionized species, which are only expected to be present if there is an incident hard radiation field. At low metallicity, stellar photoionization can power strong emission from species with ionization energies up to 54 eV (i.e., CIV, He II). Strong nebular emission from species with higher ionization energies ([Ne IV], [Ne V] and NV) is instead likely a signpost of non-stellar photoionization (i.e., AGN, fast radiative shocks).

Nearly a decade ago, the first deep measurements of rest-UV spectra were obtained at $z\gtrsim 6$. Nebular CIV emission was identified in two of the first  galaxies observed \citep{Stark2015, Mainali2017,Schmidt2017}, with the measurements revealing CIV EWs ($>$20~\AA) over an order of magnitude larger than what is commonly seen in young metal poor galaxies at lower redshifts \citep[e.g.,][]{Senchyna2019, Berg2019}. Other ground-based spectra revealed intense emission from the CIII] doublet, also with EWs well in excess of what is common at lower redshifts \citep[e.g.,][]{Stark2015, Stark2017, Hutchison2019}. These results provided our first clue that the 
radiation field (and gas conditions) of $z\gtrsim 6$ galaxies is likely significantly different from that in galaxies at $z\simeq 2$.

\begin{figure}[t]
\centering
\includegraphics[width=.9\textwidth]{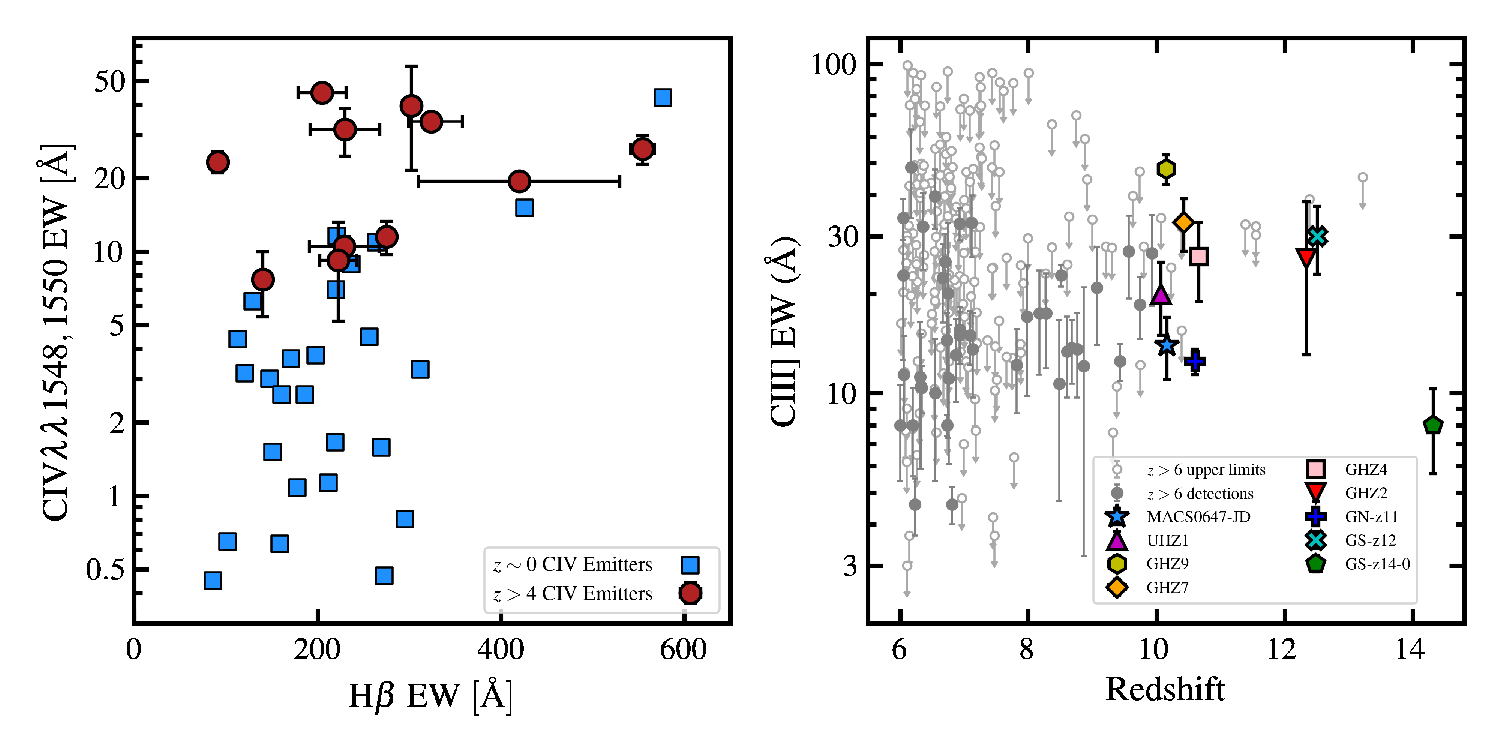}
\caption{(Left): Evidence for hard radiation fields at high redshift. Red circles show {\it JWST} measurements at $z\gtrsim 4$. At fixed H$\beta$ EW, $z\gtrsim 4$ galaxies show stronger CIV emission than metal poor nearby galaxies (blue squares), potentially indicating differences in the massive star populations in the dense clusters  at $z\gtrsim 4$.
(Right): CIII] EWs measured from \textit{JWST}/NIRSpec spectra at $z>6$. Detections at $z>10$ are highlighted: \citet{Bunker2023,Carniani2024,Castellano2024,DEugenio2024,Goulding2023,Hsiao2024,Napolitano2024b}. Extremely strong CIII] detections (EW $>10-20\ {\rm \AA}$) appear commonly at $z\gtrsim 10$. }
\label{fig:civew}
\end{figure}

{\it JWST} has begun to provide more insight into the properties of the ionizing sources. The high ionization UV lines that were seen in a handful of galaxies from the ground are now being detected in larger numbers with NIRSpec  \citep[e.g.,][]{Tang2023,Topping2024c,Castellano2024,Witstok2024a,Napolitano2024b}.  The CIV  and CIII] EWs in many of the $z\gtrsim 6$ galaxies appear in excess of what is seen in lower redshift galaxies at similar H$\beta$ EW (Fig.~\ref{fig:civew}). This suggests that $z\gtrsim 6$ galaxies may power a significantly harder radiation field at fixed stellar population age, as might be expected if stellar metallicity is lower, or the initial mass function is different. 
The current spectroscopic database is starting to give insight into the properties of galaxies that are able to power high ionization lines. 
The data suggests  CIV is a relatively common feature in galaxies with the largest sSFR. Among systems with [OIII]+H$\beta$ EW $>$ 2000~\AA, nebular CIV emission is seen in 30\% of galaxies \citep{Topping2024c}. When looking in the entire population, large EW CIV emission is less common, with only 8\% of sufficiently deep spectra having CIV  EW$>$10~\AA\ \citep{Topping2024c}. This indicates that the ionizing sources formed in the strongest bursts at $z\gtrsim 6$ produce very hard spectra with ample 48 eV photons.  The origin of the hard photons remains in question. In many galaxies, the rest-UV line ratios appear consistent with expectations for stellar photoionization \citep[e.g.,][]{Bunker2023, Topping2024a, Topping2024c,Curti2024b}, although partial AGN contribution cannot be ruled out.  But the results nevertheless suggest that the stellar populations formed in strong bursts at $z>6$ are unlike those at $z\simeq 2-3$. 
Direct measurements of stellar wind and photospheric features with moderate or high resolution {\it JWST} spectroscopy will ultimately be required for more detailed insight.  

At $z\gtrsim 10$, we rely heavily on the rest-frame UV emission features for insight, as many of the strongest rest-frame optical emission lines are redshifted out of the NIRSpec spectral window. While MIRI is starting to provide important constraints on a small number of $z>10$ galaxies \citep[e.g.,][]{Hsiao2024, Zavala2024}, it is with NIRSpec where we will get the most information. Perhaps surprisingly, four of the first bright $z>10$ galaxies observed (GNz11, GHZ2, GHZ4, GHZ7) exhibit extreme UV emission lines, with the hard ionizing radiation field, high electron densities, and elevated N/O ratios described in previous sections \citep{Bunker2023,Castellano2024,Napolitano2024b}. This may suggest that whatever physical process is responsible for the unexpected discovery of bright $z>10$ galaxies (see \S\ref{sec:lf}) is physically linked to whatever  is driving such extreme emission spectra (i.e., strong bursts, AGN).

Searches for emission lines probing ionization energies above the He$^+$-ionizing edge are also of great interest.
Nearby metal poor star forming galaxies have long been known to exhibit high ionization emission lines, likely reflecting fast radiative shocks or accretion onto high mass X-ray binaries \citep[e.g.,][]{Izotov2004}. AGN are also a plausible source of high ionization emission at high redshifts.
Meaningful  constraints on the high ionization features require very deep rest-UV spectroscopy, capable of detecting relatively weak (EW=5-10~\AA) lines.  While most surveys are currently too shallow (1-2 hrs) in the 1-2$\mu$m (observed frame) regime required to probe the rest-frame UV at $z\gtrsim 5$, we are starting to see progress in the deepest spectra. One of the first  examples is a detection of [Ne V] (and He II) emission in a seemingly-typical $z=5.59$ galaxy \citep{Chisholm2024}. This points to the presence of 97 eV photons, an energy regime not expected from stars. The [Ne V] EW in this galaxy (11~\AA) is 4 times larger than the values typically seen in metal poor star forming galaxies (where shocks may be powering the line emission), but is consistent with the EW range (6-21~\AA) typically seen in $z\simeq 1$ AGN.  Several other similar high ionization features have been reported in other studies \citep{Scholtz2023,Maiolino2024}.
We will briefly comment on the broader AGN context in \S\ref{sec:AGN}. To robustly quantify the fraction of the population with [Ne IV] or [Ne V], we will need to reach EW limits of $\simeq 5$~\AA\ ($>5-10\times$ deeper than EW limits that are common now) for a large sample of high redshift galaxies.

\section{Galaxy Clustering and Galaxy-Halo Connection}\label{sec:clustering}

The clustering of galaxies provides one of our best observational tools to constrain the dark matter halo masses that are linked to the galaxies observed in imaging surveys \citep[see][]{Wechsler2018}.
Prior to {\it JWST}, clustering measurements were conducted with the wide-area {\it Subaru} imaging and {\it HST} imaging across multiple deep fields. Measurement of the angular two-point correlation function constrained both the correlation length, r$_0$, and galaxy bias, b, of $z\simeq 4-6$ galaxies between M$_{\rm{UV}}=-22$ and $-19$ \citep[e.g.,][]{Lee2006, Harikane2016, Harikane2022, Qiu2018, Dalmasso2024a}. These studies have tended to show larger bias in more luminous galaxies, as expected if the halo masses are larger among brighter galaxies at a given redshift. Current $z\simeq 6$ constraints suggest that typical halo masses range from M$_{\rm{halo}}\sim$10$^{12}$ M$_\odot$ at  M$_{\rm{UV}}=-22$ to M$_{\rm{halo}}\sim$10$^{11}$ M$_\odot$ at  M$_{\rm{UV}}=-19$ \citep{Harikane2022}.
{\it JWST} is extending clustering analysis  to fainter and higher redshift galaxies. \citet{Dalmasso2024b} present measurement of the two-point correlation function in deep NIRCam imaging of a single field. The sample includes 7000 photometrically-selected galaxies  over $5<z<11$, with absolute magnitudes as faint as M$_{\rm{UV}}=-16$. The measurements suggest a large bias at $z\simeq 10.6$, indicating the population with M$_{\rm{UV}}<-17.3$ is hosted by dark matter halos with typical masses of M$_{\rm{halo}}\sim10^{10.6}$ M$_\odot$. \citet{Gelli2024} have shown that this bias may be larger than expected if UV scatter is the primary factor responsible for the excess in the UVLF at $z\gtrsim 12$.
Larger samples in more fields will allow these initial bias estimates to be improved in the future.

Once clustering measurements establish typical dark matter halo masses as a function of M$_{\rm{UV}}$, it becomes possible to explore the ratio of the dust-corrected SFR and halo mass. Using relations between halo mass and dark matter accretion rate \citep[e.g.,][]{Behroozi2015}, the average SFE can then be estimated as a function of halo mass and redshift. \citet{Harikane2022} presented constraints on SFR/$\dot{\rm{M}}_{\rm{halo}}$ based on a large clustering analysis, finding that the data support a near-constant SFE  at $z\simeq 5-7$ (as introduced in \S3.1). Over a wider redshift range ($z\simeq 2-7$), the evolution in the SFR/$\dot{\rm{M}}_{\rm{halo}}$ vs. M$_{\rm{halo}}$ relationship is found to be less than 0.3 dex. The relationship peaks between M$_{\rm{halo}}=10^{11}$  and 10$^{12}$ with values of  SFR/$\dot{\rm{M}}_{\rm{halo}} \sim 0.02-0.06$ at $z\simeq 6-7$ \citep{Harikane2022}. Extension of these measurements to lower halo masses will be required to better understand the UVLF at $z\gtrsim 10$ \citep{Feldmann2024}.
The galaxy-halo connection can additionally be constrained via the stellar-to-halo-mass relation (SHMR), derived from the stellar mass function using abundance matching techniques. Recent {\it JWST} studies have begun to put the first constraints on the stellar mass function at high redshifts \citep{Weibel2024, Shuntov2024, Harvey2024}, building on earlier work with {\it Spitzer} \citep{Stefanon2021}.  Here the observational picture has yet to converge. Some have argued that the SHMR remains unchanged at $z\gtrsim 6$ \citep[e.g.,][]{Stefanon2021}, as might be expected if the SFE is slowly varying. Others have argued for strong redshift evolution, with very large values indicating very efficient star formation at $z\simeq 10$ \citep[e.g.,][]{Shuntov2024}. Future work is required to explore and understand these differences.

\section{AGN in Early Galaxies} \label{sec:AGN}

It has long remained unclear observationally how many of the $z>6$ galaxies identified within deep fields harbor AGN.
For much of the past two decades, our census of $z>6$ AGN was almost solely restricted to extremely luminous quasars ($-30 < M_\mathrm{UV} < -25$), objects powered by very massive black holes ($M_\mathrm{BH}\sim10^9\,M_\odot$) that are accreting near the Eddington limit and are unobscured along our line of sight (see \citealt{Fan2023} for a review).
Because these quasars are exceptionally rare (1 per $\sim$100 deg$^2$), they are clearly not representative of the AGN population that might reside within HST-selected $z>6$ galaxies.
By the early 2020s, models and indirect empirical evidence suggested that obscured AGN should be far more common than their unobscured counterparts in the early Universe \citep{Vito2018,Davies2019,Trebitsch2019,Ni2020,Eilers2021,Gilli2022,Lupi2022,Satyavolu2023}, but these early obscured AGN were much more challenging to identify than quasars.
While broad-line CIV emission could be easily detected from the most luminous $z>6$ quasars, obscured AGN would neither be bright in the rest-frame UV continuum nor show broad line emission.
Identifying obscured AGN instead requires diagnostics from, e.g., high-ionization narrow line emission in the rest-UV, the continuum slope in the rest-frame mid-infrared, or the luminosity of radio continuum emission (see \citealt{Hickox2018} for a review).

In the year proceeding the arrival of {\it JWST} data, two independent discoveries provided the first direct empirical evidence that obscured (or at least heavily-reddened) AGN were indeed much more common than unobscured quasars at $z>6$.
\citet{Fujimoto2022} reported NOEMA confirmation of a very UV-bright ($M_\mathrm{UV} \approx -23$), yet very red Lyman-break selected object at $z_\mathrm{spec}=7.19$ within the $\approx$0.05 deg$^2$ GOODS-N field. 
This object showed a clear \textit{Spitzer}/MIPS 24$\mu$m detection implying strong hot dust emission from an AGN torus, as well as a rest-UV spectrum (from \textit{Hubble} grism data) consistent with a very red quasar template. 
A few months later, \citet{Endsley2023_radioAGN} reported ALMA confirmation of a $z_\mathrm{spec}=6.85$ object within the 1.5 deg$^2$ COSMOS field with detections spanning the near-infrared to radio regime \citep{Endsley2022_radioAGNcandidate}.
The multi-wavelength SED of this object indicated a heavily-obscured radio-loud AGN residing within a massive host galaxy undergoing an extreme dust-obscured starburst event. 
While only two objects, the \citet{Fujimoto2022} and \citet{Endsley2023_radioAGN} discoveries implied number densities of obscured/heavily-reddened AGN that are $\approx$1000$\times$ higher than classical $z\sim7$ quasars (at fixed AGN bolometric luminosity).
This begged the question of whether obscured and heavily-reddened AGN are in fact fairly common among early galaxies.
If so, we would expect to identify several more such AGN with {\it JWST} given its much redder wavelength coverage compared to \textit{Hubble}, and its unique sensitivity to broad rest-optical line emission as well as high-ionization narrow-line emission at $z\gtrsim6$.

Within the first few months of {\it JWST} science operations, several $z\sim5-9$ candidates were identified with extremely compact morphologies and peculiar photometric colors (relatively blue at $\sim$1--3$\mu$m and very red at $\sim$3--5$\mu$m) that could not be easily explained with star-forming models, raising the question of whether these systems hosted AGN \citep[e.g.,][]{Akins2023,Endsley2023,Furtak2023,Onoue2023}. 
Over the subsequent year, several such systems (nicknamed `Little Red Dots') were confirmed to show very broad (FWHM$\sim$1500--4000 km/s) H$\alpha$ or H$\beta$ emission lines implying they often fall into the class of broad-line AGN (BLAGN), albeit heavily reddened \citep[e.g.,][]{Harikane2023_BLAGN,Kocevski2023,Kokorev2023,Furtak2024,Greene2024,Lin2024,Matthee2024}. 
Current estimates suggest that these red BLAGN reside in $\sim$1\% of the UV-faint ($-20 \lesssim M_\mathrm{UV} \lesssim -18$) galaxy population at $z\sim5-8$, and perhaps $\sim$50\% of the very UV-bright ($M_\mathrm{UV} \sim -22$) population though statistics remain limited in the very luminous regime \citep{Harikane2023_BLAGN,Greene2024,Kocevski2024,Kokorev2024,Matthee2024,Taylor2024}.

The exact nature of these red BLAGN remains debated given their lack of strong mid-infrared torus dust emission and coronal X-ray emission (e.g., \citealt{Lambrides2024,Maiolino2024,PerezGonzalez2024,Williams2024}). 
Several works have estimated their bolometric luminosities by applying local calibrations based on classical type 1 BLAGN, inferring bolometric luminosities of $L_\mathrm{bol}\sim10^{44-46}$ erg/s \citep[e.g.,][]{Kokorev2023,Akins2024_UVlines,Furtak2024,Greene2024,Lin2024,Matthee2024,Taylor2024}, comparable to that of very UV-luminous quasars found over much wider areas. 
The local calibrations (again derived from classical type 1 AGN) also typically imply relatively low Eddington ratios ($\lesssim$50\%) and fairly high inferred black hole masses ($\sim10^{7-9}\,M_\odot$) for this peculiar yet abundant red BLAGN population now being discovered en masse at high redshifts. 
If these inferred black hole masses are taken at face value, they imply black hole to stellar mass ratios as large as $M_\mathrm{BH}/M_\ast \sim 0.1$, far higher than that typical of local type 1 AGN ($\sim$0.0003; \citealt{Reines2015}), which may be explained by the efficient formation of `heavy' direct collapse black hole seeds in the early Universe \citep[e.g.,][]{Volonteri2010,Natarajan2011,Natarajan2024,Pacucci2023}.
However, substantial systematic uncertainties persist in not only the black hole masses of the abundant red BLAGN at $z\gtrsim4$, but also their host stellar masses given the unclear physical origin of their red rest-optical continua \citep[e.g.,][]{Inayoshi2024,Wang2024}.
Several of the more classical unobscured type 1 AGN at $z>4$ being uncovered with {\it JWST} are also generally exhibiting evidence for very high $M_\mathrm{BH}/M_\ast$ ratios \citep[e.g.,][]{Harikane2023_BLAGN,Maiolino2023_BLAGN}, perhaps offering clearer evidence for the presence of heavy seeds.

The apparent high black hole masses inferred among the high-redshift BLAGN found with {\it JWST} may be implying that the growth of black holes often outpaced that of their host galaxies within the first billion years.
However, it has also been proposed that the high $M_\mathrm{BH}/M_\ast$ ratios in \textit{known} BLAGN is at least in part a selection bias, where the most obvious AGN found over small areas will be low-mass galaxies with up-scattered black hole masses resulting in fairly high AGN to stellar luminosity ratios (\citealt{Li2024}; c.f., \citealt{Pacucci2023}).
Moreover, given that the multi-wavelength (X-ray, mid-infrared, far-infrared, and radio) properties of these high-redshift BLAGN generally do not match expectations from local type 1 AGN \citep[e.g.,][]{Akins2024_CWeb,Lambrides2024,Maiolino2024}, it is unclear whether local bolometric luminosity and black hole mass calibrations are appropriate for this population. 
It has been proposed that many of the observed properties of \textit{JWST} BLAGN can be explained by models with super-Eddington accretion \citep[e.g.,][]{Lambrides2024,Lupi2024,King2024,Pacucci2024,Volonteri2024}, which may bring down the estimated black hole masses by an order of magnitude if not more. 
In some super-Eddington accretion models, relatively weak high-ionization lines in the rest-frame UV may be expected, which current (albeit limited) data do support \citep{Akins2024_UVlines,Lambrides2024}.
The weak X-ray emission from \textit{JWST} BLAGN may also be due to a high covering fraction of Compton-thick gas (hydrogen column density $N_H > 1.5\times 10^{24}$ cm$^{-2}$), likely mostly in the broad line region \citep[e.g.,][]{Maiolino2024}. This picture is supported by the growing number of Balmer absorption features being detected in higher resolution rest-optical spectra.
The absorbing gas would  need to have very low dust-to-gas ratios to not fully attenuate the rest-optical broad-line emission.
Further multi-wavelength {\it JWST} follow-up is required for to better establishe the nature of newly-discovered high-redshift BLAGN population. 

Obscured AGN are also beginning to be identified with {\it JWST} data, albeit not as readily as the BLAGN described above. In the most clear example (introduced in \S9), the extremely high-ionization ($>$97 eV) [NeV]$\lambda$3427 line has been  detected (7$\sigma$) in a moderately-massive ($M_\ast \approx 10^8\,M_\odot$) galaxy at $z_\mathrm{spec} = 5.59$ \citep{Chisholm2024}, with low-to-high ionization line ratios implying photoionization from both massive stars and an intermediate-mass black hole ($\sim10^6\,M_\odot$).
Other (sometimes tentative) high-ionization line detections have been reported among $z\gtrsim6$ galaxies \citep[e.g.,][]{Fujimoto2023,Scholtz2023,Treiber2024}, with line ratios also consistent with at least partial AGN photoionization.
As noted above, the BLAGN population appear to generally show weak high-ionization UV lines that are potentially consistent with expectations from super-Eddington accretion models \citep{Lambrides2024} or large covering factor of the BLR  \citep{Maiolino2024}, so it remains unclear whether this is an efficient selection method for obscured high-redshift AGN.
Several candidate obscured AGN at $z\gtrsim6$ are also now being identified from strong rest-frame mid-infrared emission, presumably from a dusty torus illuminated by the hot accretion disk \citep{Yang2023,Lyu2024}. 
Further {\it JWST} surveys are ongoing to better quantify the abundance of obscured, mid-IR bright AGN at high redshifts.

\begin{figure}[t]
\centering
\includegraphics[width=1\textwidth]{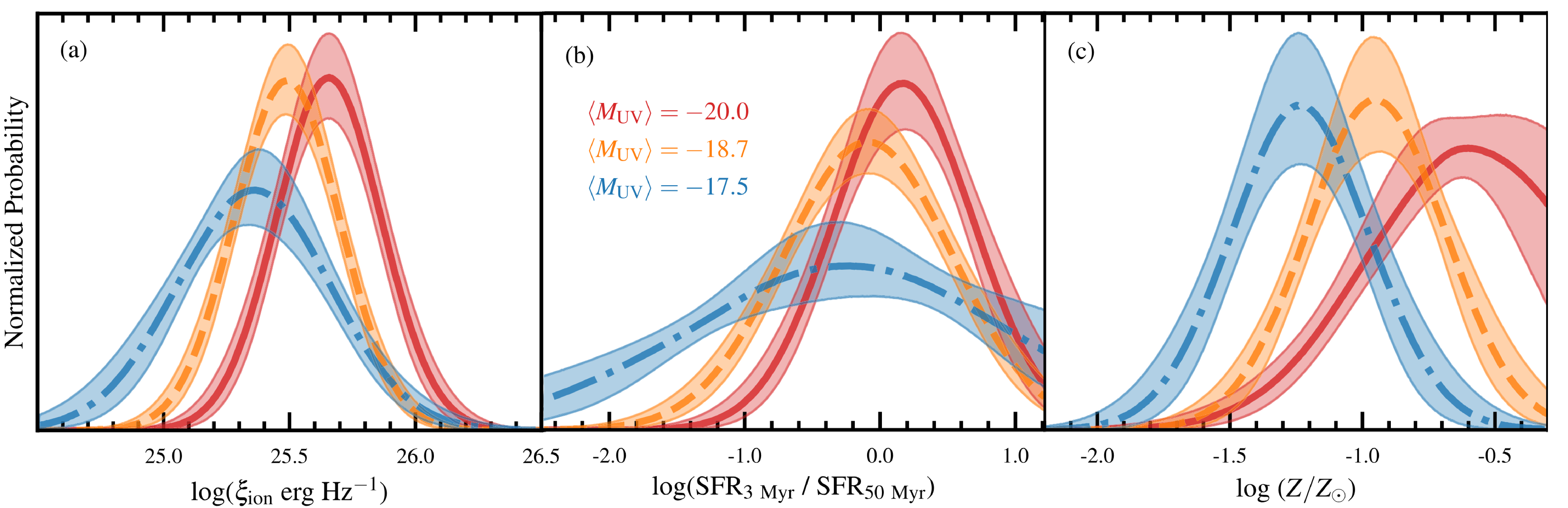}
\caption{(Left:) Ionizing photon production efficiency ($\xi_{\rm{ion}}$) distributions in three different absolute magnitude bins at $z\sim 6$, as derived from SEDs fit with flexible SFHs. The distribution in $\xi_{\rm{ion}}$ is broad, reflecting the on and off modes of bursty star formation.  (Middle:) Distribution of $\rm SFR_{3Myr}/SFR_{50Myr}$ recovered in SED models. The faintest galaxies are more likely to be in an SFR lull with low   $\xi_{\rm{ion}}$ than more luminous galaxies.
(Right:) Distributions of metallicity recovered simultaneously in the modeling. These properties were derived using SED models that assume a two-component SFH and $f_{\rm esc}=0$. Plots are 
adapted from \citet{Endsley2024}. }
\label{fig:xiion}
\end{figure}

\section{Outflows in Absorption and Emission}\label{sec:outflows}

At lower redshifts, signs of large-scale outflows are ubiquitous in star forming galaxies, both through interstellar absorption lines \citep[e.g.,][]{Shapley2003, Steidel2016, Du2016, Calabro2022} and broad rest-optical emission lines \citep[e.g.,][]{Shapiro2009, Newman2012, ForsterSchreiber2019,Davies2019ha,Kehoe2024}. Given the large star formation rate surface densities at $z\gtrsim 5$ (\S\ref{sec:sizes}), we may expect to commonly see similar signs of outflows in $z\gtrsim 6$ galaxies. Prior to {\it JWST}, 
direct constraints on feedback at $z\gtrsim 6$ were out of reach observationally. The first look at rest-optical emission lines at $z\gtrsim 6$ clearly revealed broad components in the forbidden [OIII] line (100-800 km/s) in many continuum-selected galaxies. The incidence fraction of broad outflows has been found to be between 17\% and 30\% \citep{Xu2023, Zhang2024, Carniani2024}. This fraction will be set by the duty cycle and opening angle of the outflows. Current work suggests that broad wings may be more common at $z\simeq 3-9$ than they are at $z\simeq 1$ \citep{Xu2023}. Estimates of of the mass loading factor ($\eta=\dot{M}_{out}/$SFR) are varied, with some arguing for similar values as found at lower redshift ($\eta=0.1-1$, \citealt{Xu2023}) and others suggesting higher values ($\eta=2$, \citealt{Carniani2024}).

With NIRSpec it is now also possible to identify and characterize absorption line kinematics. Because deep rest-UV spectroscopy in the required medium or high resolution settings remains rare, current absorption line samples are still small. Nevertheless, existing measurements have been made in two bright galaxies, with blueshifted low and high ionization absorption lines clearly indicating outflowing gas with speeds of $\simeq 200$ km/s \citep{Topping2024b,Topping2024c}.
The EWs of the absorption lines range from 0.5 to 2.0~\AA, similar to those seen in spectra of similar galaxies at lower redshifts. In the future, composites in bins of different galaxy properties should enable insight into the kinematics and covering fraction of the ISM and CGM in $z\gtrsim 7$ galaxies.

\section{Galaxies and Reionization}\label{sec:reionization}
In this section, we describe recent progress in constraining the contribution of galaxies to reionization (\S\ref{sec:galaxycontribution}) and we discuss {\it JWST} spectroscopic measurements of the prevalence of Ly$\alpha$ emission at $z\gtrsim 6$ (\S\ref{sec:lya}).

\subsection{Contribution of Galaxies to Reionization}\label{sec:galaxycontribution}

As knowledge of the census of galaxies improves, so does our ability to compute their contribution to reionization. The cosmic ionization rate produced by star forming galaxies (n$_{\rm{ion}}$) can be estimated from determinations of the UV luminosity density and the ionizing efficiency of galaxies, 
\begin{equation}
\rm{n_{ion} = f_{esc} ~ \xi_{ion} ~ \rho_{UV}}
\end{equation}
where f$_{\rm{esc}}$ is the fraction of ionizing photons which escape to the IGM, and $\xi_{\rm{ion}}$ is the ionizing production efficiency, defined as the ratio of the of hydrogen ionizing photon production rate and the emergent far UV luminosity at 1500~\AA. 
Both f$_{\rm{esc}}$ and $\xi_{\rm{ion}}$ may depend on redshift, and they may also vary with galaxy mass or luminosity. In lockstep with efforts to measure $\rho_{\rm{UV}}$ from UVLFs, considerable attention has also focused on measurement of the ionizing output of high redshift galaxies. We briefly summarize recent results below.

The ionizing photon production efficiency is  constrained in several of ways. The more empirical approach uses direct measurement of the luminosity of a hydrogen recombination line (i.e., H$\alpha$, H$\beta$) to  estimate  the ionizing photon production rate under assumption of case B recombination. This is then combined with a dust-corrected measurement of the UV continuum luminosity to arrive at $\xi_{\rm{ion}}$. The  more model-based approach involves fitting observations 
(SEDs, spectra) with spectral modeling tools, constraining the range of model ionizing photon production efficiencies that reproduce the data. 
It has been shown at lower redshifts that the ionizing photon production efficiency is  correlated with the H$\beta$ and [OIII] EW \citep{Chevallard2018,Tang2019}. Physically this result follows naturally from the relationship between $\xi_{\rm{ion}}$ and stellar population age. Assuming a constant star formation history, the UV continuum luminosity will increase steadily for 100 Myr before it reaches a steady state, whereas the ionizing output will equilibrate after roughly 10 Myr. The ionizing photon production efficiency will therefore be largest at very young ages and large H$\beta$ and [OIII] EWs.

The first measurements of $\xi_{\rm{ion}}$ at $z\gtrsim 6$ with {\it JWST} demonstrated that early galaxies mostly follow a similar relationship with [OIII] and H$\beta$ EWs, with average values increasing from $\xi_{\rm{ion}}$ =10$^{25.1}$~erg$^{-1}$~Hz at [OIII]+H$\beta$ EW=300~\AA\ to $\xi_{\rm{ion}} = 10^{25.9}$~erg$^{-1}$~Hz at [OIII]+H$\beta$ EW=3000~\AA\ (e.g., \citealt{Tang2023}). Because the average sSFR (and hence [OIII]+H$\beta$ EW) increases toward higher redshift (\S\ref{sec:galaxyprops}), the average ionizing photon production efficiency will also increase at earlier epochs. This was predicted and demonstrated prior to {\it JWST} \citep{Tang2019,Endsley2021} and has now been shown to be true with {\it JWST} spectroscopy. Among moderately luminous (M$_{\rm{UV}}<-19$) $z\simeq 6-8$ galaxies the average ionizing photon production efficiency is $\xi_{\rm{ion}}$ =10$^{25.5-25.7}$~erg$^{-1}$~Hz (\citealt{Prieto-Lyon2023,Endsley2024,Pahl2024,Simmonds2024,Saxena2024}). Similarly large ionizing photon production efficiencies have been found among extremely faint galaxies identified via gravitational lensing \citep{Atek2024}. However it is now clear that the strong rest-optical lines are not truly ubiquitous at very high redshift. As we described in \S\ref{sec:galaxyprops}, there is a sizable population of galaxies with weak rest-optical emission lines, 
plausibly corresponding to systems that have 
experienced a recent decline in star formation. These galaxies will have a deficit of O stars and correspondingly lower $\xi_{\rm{ion}}$ values.\footnote{Identification of weak [OIII] emission is not sufficient to confirm low ionizing photon production efficiency, as low gas-phase metallicity will also act to depress [OIII] EWs. 
The $\xi_{\rm{ion}}$ relations described above have been derived at redshifts where both H$\alpha$ and [OIII] can be constrained, allowing the degeneracy between $\xi_{\rm{ion}}$ and metallicity to be broken.
Because of this degeneracy, care must be taken when using low redshift empirical relations between [OIII] EW and $\xi_{\rm{ion}}$.} The discovery of this sub-population suggests that the 
ionizing photon production efficiency distribution may have much larger scatter than previously thought. 
During SFR lulls, galaxies will be found with very low values ($\xi_{\rm{ion}} \lesssim 10^{25.0-25.2}$~erg$^{-1}$~Hz),\footnote{When $\xi_{\rm{ion}}$ is derived from SEDs, low ionizing photon production efficiencies associated with SFR lulls will only be obtained if sufficient SFH flexibility is allowed in the modeling, and if constraints on flux excesses from both [OIII] and H$\alpha$ are available from medium-band photometry.  }  and during strong bursts galaxies will appear at the opposite side of the distribution ($\xi_{\rm{ion}} \sim 10^{25.7-26.0}$~erg$^{-1}$~Hz).
Since galaxies with recent SFR declines appear to predominantly be at the faint end of the luminosity function (M$_{\rm{UV}}=-17.5$; \citealt{Endsley2024b}), 
the low $\xi_{\rm{ion}}$ tail is preferentially found among low luminosity galaxies. Luminosity-dependent $\xi_{\rm{ion}}$ distributions  have now been derived \citep{Endsley2024} and are being integrated into  basic calculations of the contribution of galaxies to reionization (Fig.~11).\footnote{It is worth re-emphasizing that current data cannot rule out ionizing photon leakage as the root cause of weak emission line spectra. In this very different picture, the ionizing photon production efficiencies would not be reduced, and  the escape fractions would be very large. However many galaxies with weak lines have UV slopes with values  ($\beta>-2$) that do not appear likely to host such extremely large ionizing photon escape fractions. SFR variability appears to provide a more likely explanation in these cases. However deep moderate resolution spectroscopy of weak line galaxies is required to verify this picture. } 

Considerable effort have also been placed in quantifying the range of   escape fractions which are likely to be common at $z\gtrsim 5$. Since the opacity of the IGM prohibits direct constraints on the escape fraction at $z\gtrsim 4$, efforts instead mostly focus on indirect indicators of ionizing photon leakage. Over the past decade, there has been much focus in establishing the range of properties which are correlated with large escape fractions at $z\simeq 0-3$ \citep[e.g.,][]{Reddy2016, Steidel2018, Flury2022, Chisholm2022, Begley2022, Pahl2022, Pahl2023, Jaskot2024a, Jaskot2024b}. While results vary somewhat from study to study, it has generally been found that escape fractions are largest in galaxies with low masses, blue UV colors, large Ly$\alpha$ EWs, and small Ly$\alpha$ peak separations. Some studies have further shown escape fractions are more likely to be large in galaxies with extremely large H$\beta$ EWs and large O32 values. 
Investigations with {\it JWST} have attempted to quantify the escape fractions based on empirical relationships between f$_{\rm{esc}}$ and galaxy properties which can be measured (i.e., UV slope, M$_\star$, O32, H$\beta$ EW, $\Sigma_{\rm{SFR}}$).\footnote{Ly$\alpha$ is significantly impacted by the IGM at very high redshift, making it less useful as a LyC indicator at $z\gtrsim 5$. Even at $z\simeq 5-6$, centrally-peaked Ly$\alpha$ profiles (indicating low HI density channels that may facilitate ionizing photon escape) will have their line profiles significantly modulated by the residual HI fraction in the IGM.
} These results tend to indicate that galaxies at $z\gtrsim 6$ will have properties which  suggest an average escape fraction of $\sim 0.1$ \citep[e.g.,][]{Gazagnes2024, Mascia2024, Saxena2024}. Since high redshift galaxies appear to undergo significant variability in their SFR, it is plausible that they also have time-variable escape fractions, with short windows where very large fractions  ($\gtrsim 0.5$) of ionizing radiation is able to escape \citep[e.g.,][]{Trebitsch2017,Barrow2020,Ma2020,Flury2024}. Such galaxies would appear like the Sunburst arc, a well-known strong leaker at $z=2.38$ \citep{Rivera-Thorsen2017}. When escape fractions are near-unity, the nebular emission will be significantly weakened, altering the SED. The UV  colors can approach the intrinsic stellar continuum colors, which we expect to be close to $\beta\simeq -3$ for dust-free galaxies with large escape fractions (as is seen in the Sunburst arc, \citealt{Kim2023}). Emission lines will also be weakened, although in many cases the intrinsic equivalent widths are so large that galaxies remain large EW rest-optical line emitters even if they leak 50\% of their ionizing radiation. As noted in \S\ref{sec:uvslopes}, there is a subset of $z\gtrsim 5$ galaxies which show UV colors approaching $\beta\sim -3$ \citep{Topping2022,Topping2024a,Cullen2024, Austin2024,Hainline2024}. These galaxies are potentially very similar to the Sunburst arc, leaking large fractions ($\gtrsim 0.3-0.5$) of their ionizing radiation. \citep{Topping2024a} demonstrate that galaxies with $\beta<-2.8$ also systematically have weaker flux excesses associated with emission lines, suggesting that these systems face an overall reduction in nebular emission. 
Spectroscopy is required to explore whether this population is truly leaking large escape fractions (in which case we may expect to see stellar P-Cygni features in deep continuum spectra) or whether they instead have reduced nebular emission following a burst of star formation (in which case there should not be strong O star features). 

Equipped with new measurements of ionizing output of early galaxies,  recent studies have begun to investigate the range of plausible reionization histories consistent with existing data. \citet{Gelli2024} demonstrate that because of the excess UV photon density at $z\gtrsim 10$, it is possible that reionization may be underway somewhat earlier than previously predicted in some models. They consider reionization models that include UV scatter that  reproduces the UVLF, and they also adopt the M$_{\rm{UV}}$-dependent ionizing photon production efficiencies described above. 
The results show that in cases where the escape fraction is very large during bursts, the IGM may already be 10-20\% ionized at $z\simeq 10$.  Without the inclusion of UV scatter, the ionized fraction would be much lower at $z\gtrsim 10$ (see also \citealt{Furlanetto2022, Nikolic2024}). The wide dispersion in the ionizing photon production efficiencies is  found to be important in avoiding scenarios where reionization ends early and violates Thomson scattering constraints from the CMB \citep{Munoz2024}. Whether the IGM is actually 10-20\% ionized at $z\simeq 10$ is not known. This prediction not only relies on knowledge of the escape fraction, but it also depends on the  luminosity at which the UVLF turns over. In the following subsection, we discuss {\it JWST} efforts to use spectroscopic measurements of Ly$\alpha$ emission  to probe the early stages of reionization. 

\begin{figure}[t]
\centering
\includegraphics[width=1\textwidth]{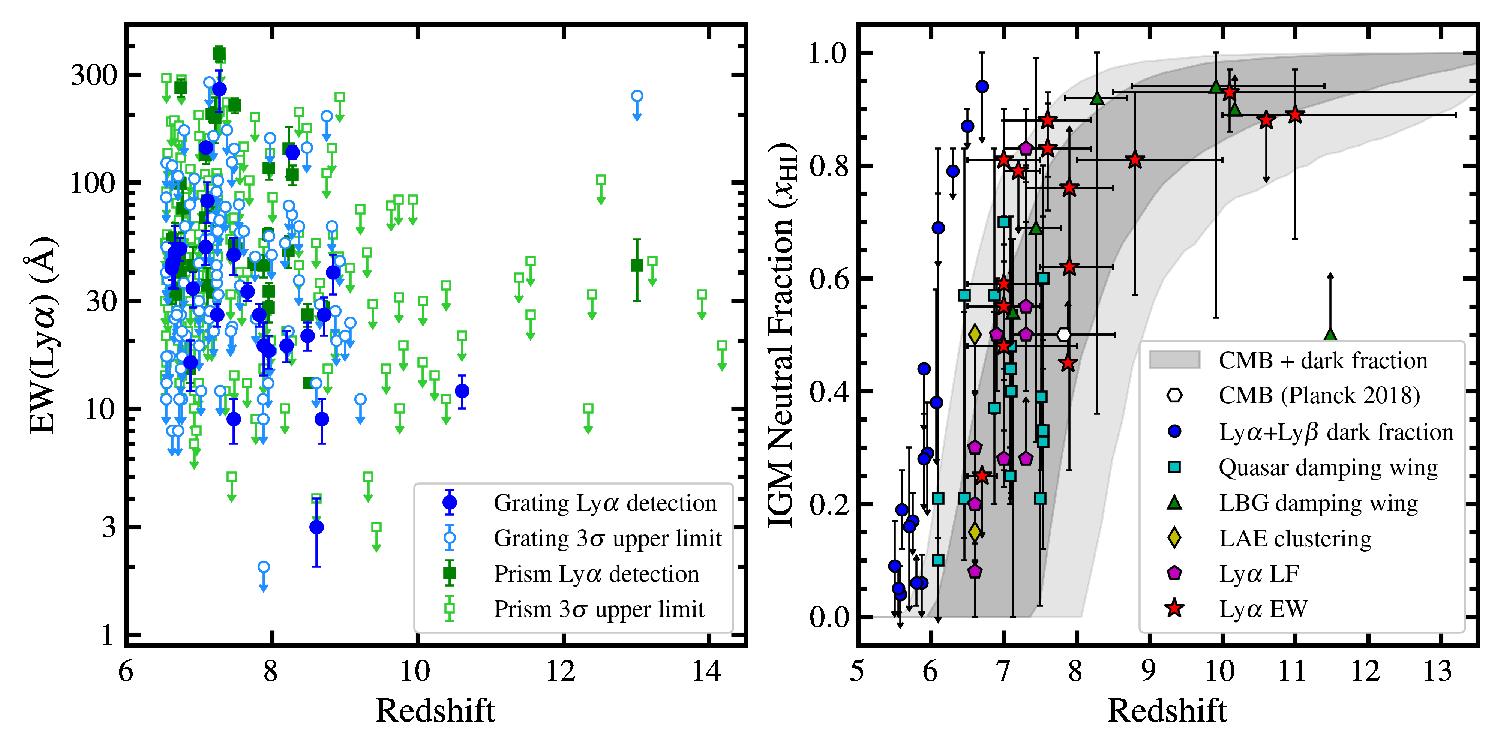}
\caption{Left: Census of Ly$\alpha$ EWs measured from \textit{JWST}/NIRSpec spectra at $z>6.5$ \citep{Carniani2024,Castellano2024,Jones2024,Tang2024c,Witstok2024b}. Very large Ly$\alpha$ EWs ($>100$~\AA) are visible at $z\simeq 6-8$ but disappear above $z\simeq 8.5$.
Right: Evolution of IGM neutral fraction ($x_{\rm HI}$) in the reionization era. The dark (light) grey shaded region is the $1\sigma$ ($2\sigma$) constraint derived from CMB optical depth and Ly$\alpha$ and Ly$\beta$ forest dark pixel fraction \citep{Mason2019b}. Individual datapoints show the IGM neutral fractions derived from multiple observational probes: CMB optical depth (white hexagon; \citealt{Planck2020}), Ly$\alpha$ and Ly$\beta$ forest dark pixel fraction (blue circles; \citealt{McGreer2015,Zhu2022,Zhu2024,Jin2023,Spina2024}), Ly$\alpha$ damping wing absorption of quasars (cyan squares; \citealt{Greig2017,Greig2019,Greig2022,Banados2018,Davies2018,Durovcikova2020,Durovcikova2024,Wang2020,Yang2020}), Ly$\alpha$ damping wing absorption of Lyman break galaxies (green triangles; \citealt{CurtisLake2023,Hsiao2024,Umeda2024}), the clustering of LAEs (yellow diamond; \citealt{Ouchi2010,Ouchi2018,Sobacchi2015}), Ly$\alpha$ luminosity function (magenta pentagon; \citealt{Ouchi2010,Konno2014,Konno2018,Zheng2017,Inoue2018,Goto2021,Morales2021,Ning2022}), and Ly$\alpha$ EW (red stars; \citealt{Mason2018,Mason2019a,Hoag2019,Whitler2020,Bolan2022,Bruton2023,Morishita2023,Jones2024,Nakane2024,Tang2024c}). Both plots are adapted from \citet{Tang2024c}. }
\label{fig:lya}
\end{figure}

\subsection{Ly$\alpha$ and Reionization}\label{sec:lya}

Over the last decade, knowledge of the timeline of the latter half of reionization has developed rapidly (Fig 12b; \citealt{Ouchi2020,Fan2023}). Deep quasar spectroscopy  reveals the process is mostly complete at $z\simeq 5.3$. At $z\simeq 6$, upwards of 5-20\% of the IGM is likely still  neutral. By $z\simeq 7$, the process is at its midpoint, with an ionized fraction approaching 50\%. Knowledge of the first half of reionization is more limited. Characterization of the Ly$\alpha$ emission line emitted by star forming galaxies  provides one of our only means of learning about the early evolution of the IGM. Owing to the strong damping wing cross section provided by neutral hydrogen, Ly$\alpha$ will be attenuated as the IGM becomes more neutral \citep{MiraldaEscude1998,Mesinger2008}. Early investigations focused on tracking the luminosity function of Ly$\alpha$ emitters identified with narrowband filters (\citealt{Malhotra2004,Hu2010,Ouchi2010}, \S2.2). In the last fifteen years, efforts have additionally focused on a complementary approach using spectroscopic measurements of the Ly$\alpha$ EW distribution in continuum-selected galaxies \citep{Stark2010,Fontana2010}. The integral of the Ly$\alpha$ EW distribution above a fixed threshold (i.e., EW = 25~\AA) gives the so-called Ly$\alpha$ fraction. As the IGM grows more neutral, the Ly$\alpha$ fraction and Ly$\alpha$ emitter number density will decrease. Over the past decade, work has been conducted to map the observed Ly$\alpha$ emitter evolution into constraints on the  reionization timeline \citep[e.g.,][]{Mason2018}. 

As continuum-selected $z\simeq 7-10$ galaxy samples emerged following the installation of WFC3/IR on {\it HST} (see \S2), it became possible to apply the Ly$\alpha$ fraction test to the first half of reionization. Ground-based spectroscopy targeted Ly$\alpha$ in several hundred galaxies between 2009 and 2022, leading to robust confirmations of Ly$\alpha$ in $\sim 18$ galaxies at $z\gtrsim 7$ (Fig 1a), mostly in fairly bright (H=25-26) systems for which ground-based spectrographs could reach useful sensitivity limits. No Ly$\alpha$ detections were obtained above $z\simeq 8.7$.  The results pointed to a declining Ly$\alpha$ fraction toward higher redshift \citep[e.g.,][]{Schenker2014,Pentericci2018,Bolan2022}, providing one of our first observational hints that the IGM was likely significantly neutral at $z\simeq 7-8$.

{\it JWST} spectroscopy has recently ushered in a new era of Ly$\alpha$ spectroscopy. The sensitivity of NIRSpec has enabled Ly$\alpha$ detections in extremely faint (H=30) galaxies at $z\gtrsim 7$ \citep[e.g.,][]{Saxena2023,Chen2024}, nearly 100$\times$ less luminous than what was achieved from the ground. The access to strong rest-optical emission lines allow redshifts to be determined whether or not Ly$\alpha$ is present, leading to far more reliable upper limits in cases where Ly$\alpha$ is not detected.   
Measurement of non-resonant emission lines enables  Ly$\alpha$ peak velocity offsets (with respect to the systemic redshift) to be determined, providing a new window on how the IGM is modulating the line profile. Detection of hydrogen Balmer lines  (H$\beta$, H$\alpha$) allow  Ly$\alpha$ escape fractions to be estimated, under basic assumptions of case B recombination. 

Current {\it JWST} spectroscopic analyses have been made on samples of over 200 galaxies spread across 4 deep fields (Fig 12a; \citealt{Jones2024,Nakane2024,Napolitano2024a,Tang2024c}). Over 30 galaxies have been identified with Ly$\alpha$ emission using NIRSpec. The results confirm the decline in the Ly$\alpha$ fraction measured prior to {\it JWST}, with the fraction of continuum-selected galaxies showing strong Ly$\alpha$ (EW$>$25~\AA) or large Ly$\alpha$ escape fractions ($>$0.2) decreasing by a factor of 4 between $z\simeq 5$ and $z\simeq 9$, indicating a rapid change in the transmission of Ly$\alpha$ photons. Only two galaxies have been identified with Ly$\alpha$ emission at $z\gtrsim 10$ \citep{Bunker2023,Witstok2024b}, suggesting a continued drop in transmission. Work is ongoing to investigate additional (non-IGM) factors that may be impacting the Ly$\alpha$ evolution. Continuum spectroscopy is making it possible to measure the statistical incidence of damped Ly$\alpha$ absorbing systems in the galaxy population as a function of redshift \citep{Heintz2024}, and characterization of Ly$\alpha$ samples at the tail end of reionization are better establishing the dependence of Ly$\alpha$ on galaxy properties \citep{Lin2024,Tang2024a}. These studies offer the prospect for significantly improving the mapping between Ly$\alpha$ fraction evolution and IGM HI fractions. Current estimates are shown in Figure 12b. While uncertainties remain significant, the Ly$\alpha$ fraction evolution is consistent with the IGM reaching over 80\% neutral at $z\gtrsim 9$. As deeper rest-UV spectroscopy is obtained in years to come, statistical uncertainties on the Ly$\alpha$ fraction should improve considerably.

While most early galaxies face significant attenuation of their  Ly$\alpha$ emission, {\it JWST} has revealed a small number of $z\gtrsim 7$ galaxies with extremely large Ly$\alpha$ EWs  ($>$100~\AA) and similarly large Ly$\alpha$ escape fractions\footnote{Some caution must be taken in interpreting Ly$\alpha$ escape fractions in the absolute sense since they depend on recombination case assumptions \citep{McClymont2024,Scarlata2024}.} and small Ly$\alpha$ velocity offsets \citep{Saxena2023,Tang2024b}. Such strong Ly$\alpha$ emission likely suggests enhanced IGM transmission, as would be expected in large ionized bubbles in the mostly neutral IGM. 
In the near future, it will be possible for {\it JWST} to begin to test this picture. The NIRCam grism allows the three-dimensional distribution of 
[OIII] emitting galaxies to be mapped across deep fields \citep[e.g.,][]{Meyer2024}, identifying $\sim 1$ physical Mpc regions that are extremely overdense where bubbles may be present. Follow-up Ly$\alpha$ spectroscopy with NIRSpec offers the potential to quantify the Ly$\alpha$ properties of galaxies in the overdensities, identifying the extent to which Ly$\alpha$ transmission is boosted. Current studies demonstrate that strong Ly$\alpha$ emitters  often appear clustered in overdensities, but the data are not yet sufficient to measure whether the Ly$\alpha$ EW distribution is enhanced within the overdense structures \citep{Tang2024c}. Future targeted spectroscopy should allow the sizes of ionized regions to be constrained with {\it JWST} in the near future \citep[e.g.,][]{Lu2024}.

\section{Summary}\label{sec:summary}%

The last few years have witnessed rapid progress in our understanding of high redshift galaxies. Deep imaging with {\it JWST} has pushed  the redshift frontier back to $z\simeq 14$, while also allowing new insights into the internal properties and sizes of high redshift galaxies. The spectroscopic capabilities of {\it JWST} have also proven transformative, opening our first window on the ionizing sources, gas conditions, and large scale structure in the first billion years.  These investigations have been complemented by continued characterization of gas and dust with ALMA. Below we briefly summarize several of the most important advances from these observations.

1. The galaxy census at $z\gtrsim 10$ has revealed a larger number density of galaxies than many expected. It has been shown that the $z\gtrsim 10$ UVLF can be reproduced via changes in the mean star formation efficiency or the UV scatter in the  M$_{\rm{UV}}$-M$_{\rm{halo}}$ relation. Further explanations include the ejection of dust and top-heavy IMFs, both of which can also boost the UV output. Additional observations will be required to identify which of these physical processes are most important in explaining the early evolution of the UVLF.

2. NIRCam SEDs have enabled characterization of stellar masses, sSFRs, and light-weighted stellar population ages for large samples of high redshift galaxies.  Typical stellar masses at $z\simeq 7$ are found to vary from  M$_{\rm{\star}}\sim10^{9.5}$ M$_\odot$ at the bright end of the luminosity function (M$_{\rm{UV}}=-22$) to M$_{\rm{\star}}\sim10^{7.0}$  M$_\odot$ at the faint end (M$_{\rm{UV}}=-17$). Galaxies at these redshifts tend to be dominated by fairly young stellar populations, although Balmer breaks do exist in some cases. The average sSFR  increases with redshift at fixed stellar mass, with a power law form that is consistent with expectations if the mean sSFR is regulated by the baryon accretion rate.

3. Observations are beginning to provide evidence that star formation histories  have significant time variability at high redshift. Some systems are observed in the midst of bursts, when the sSFR, [OIII]+H$\beta$ EWs, and $\xi_{\rm{ion}}$ are very large. Others appear likely to be observed during  SFR lulls following a recent burst. These systems have low values of sSFR,  [OIII]+H$\beta$ EWs, and $\xi_{\rm{ion}}$. Future work will  better characterize the average duty cycle and UV variability experienced by galaxies as a function of mass and redshift. 

4. Galaxies at $z\gtrsim 5$ are very compact, resulting in very high star formation rate surface densities and high stellar mass densities.  A large fraction of star formation appears to occur in dense bound star cluster complexes. When observed at very young ages ($<$5 Myr), the dense clusters show a large N/O abundance pattern that is similar to that seen in many globular cluster stars. 

5. The UV colors in $z\gtrsim 10$ galaxies are consistent with negligible dust attenuation. In contrast,   many UV luminous $z\simeq 7$ galaxies often show significant dust continuum emission with ALMA. This may suggest significant evolution in the dust content of the galaxies  at the bright end of the UVLF between $z\simeq 10$ and $z\simeq 7$.  The UV colors at $z\gtrsim 5$ also reveal a population of extremely blue galaxies ($\beta\sim-3$) which may correspond to systems with large ionizing photon escape fractions. 

6. {\it JWST} has revealed the first statistical insights into the ionized gas-phase properties of $z\gtrsim 5$ galaxies.
The mass metallicity relation is already in place at very high redshift, with typical oxygen abundances of 10\% solar metallicity at $z\gtrsim 5$. Only mild evolution is seen in the metallicities of galaxies at fixed stellar mass between $3<z<10$.  In contrast, the average ionization conditions and electron density do appear to evolve at $z\gtrsim 5$, potentially  driven by the shift toward larger sSFRs at higher redshift.

7. Relative abundance measurements from rest-UV spectroscopy have revealed a subset of metal poor galaxies with large N/O ratios, an abundance pattern that is unlike that in nearly all galaxies at lower redshifts. The stellar origin of the large N/O ratios is debated, with one of the more common explanations involving a population of very massive stars or super-massive stars.  
The C/O ratios at $z\gtrsim 5$ appear more typical of what is seen in metal poor galaxies at lower redshifts.

8. Prominent high ionization lines (CIV, He II) appear present in galaxies that are  in the midst of strong bursts. The CIV EWs often extend above values seen in nearby metal poor galaxies with similar H$\beta$ EWs, particularly in bright galaxies at $z\gtrsim 10$. The origin of the hard radiation fields is not currently known. It may point to a more extreme massive star population, possible formed in dense bursts of star formation. Alternatively it may signal contribution from  AGN photoionization.

9. An abundant population of red broad-line AGN is now being discovered at $z\sim5-9$. From local calibrations they appear to generally have very massive black holes relative to their host galaxy stellar masses. However, the nature of many of these early AGN remain debated given their peculiar multi-wavelength properties. The first statistical samples of obscured AGN at $z>5$ are also now being uncovered via very high ionization line emission (i.e. [Ne V], [Ne IV]) and strong rest-frame mid-infrared emission.

10. Constraints on the contribution of galaxies to reionization have improved with new measurements of the UV luminosity density
and ionizing efficiency of early galaxies. Models that reproduce the UVLF at $z\gtrsim 10$ suggest that reionization may start somewhat earlier than previously thought, with potentially 
10-20\% of the IGM ionized at $z\simeq 10$.
Ly$\alpha$ spectroscopy with {\it JWST} provides one avenue of testing this picture.

\begin{ack}[Acknowledgments]

The authors thank  Charlotte Mason, Zuyi Chen, and Lily Whitler for useful conversations.
\end{ack}

\seealso{Robertson, B. 2022, ARAA, 60, 121. Dayal, P. \& Ferrara, A. 2018, Phys. Rep. 780, 1.
Ouchi, M., Ono, Y. \& Shibuya, T. 2020, ARAA, 58, 617.
}

\bibliographystyle{Harvard}
\bibliography{reference}

\end{document}